\documentclass[conference]{IEEEtran}
\IEEEoverridecommandlockouts
\usepackage{fancyhdr}
\usepackage{cite}
\usepackage{amsmath,amssymb,amsfonts}
\usepackage{algorithmic}
\usepackage{graphicx}
\usepackage{textcomp}
\usepackage{xcolor}
\usepackage{array}    
\usepackage{fancyhdr}
\usepackage{graphicx}
\usepackage{amsmath}
\usepackage{array}
\usepackage{amsfonts}
\usepackage{multirow}
\usepackage{booktabs}

\def\BibTeX{{\rm B\kern-.05em{\sc i\kern-.025em b}\kern-.08em
    T\kern-.1667em\lower.7ex\hbox{E}\kern-.125emX}}

\usepackage{titlesec}

\usepackage{amsmath}

\renewcommand{\thesection}{\arabic{section}}
\renewcommand{\thesubsection}{\arabic{section}.\arabic{subsection}}
\renewcommand{\thesubsubsection}{\arabic{section}.\arabic{subsection}.\arabic{subsubsection}}

\titleformat{\section}[block]{\bfseries\normalsize}{\thesection\quad}{0pt}{}  
\titleformat{\subsection}[block]{\itshape\normalsize}{\thesubsection\quad}{0pt}{}  
\titleformat{\subsubsection}[block]{\itshape\normalsize}{\thesubsubsection\quad}{0pt}{} 

\begin{document}
\rmfamily

\title{{\fontsize{16}{19}\selectfont \textbf{Teager Energy Operator as a Metric to Evaluate Local Synchronization of Power System Devices}}}

\author{

    \IEEEauthorblockN{{\fontsize{12}{14}\selectfont Bruno Pinheiro}}
    \IEEEauthorblockA{
    \textit{\fontsize{10}{12}\selectfont Universidade Estadual de Campinas}\\
    \fontsize{10}{12}\selectfont Campinas, Brazil \\
    \fontsize{10}{12}\selectfont b229989@dac.unicamp.br}
    \and
   
    \IEEEauthorblockN{{\fontsize{12}{14}\selectfont Ignacio Ponce}}
    \IEEEauthorblockA{
    \textit{\fontsize{10}{12}\selectfont University College Dublin}\\
    \fontsize{10}{12}\selectfont Dublin, Ireland \\
    \fontsize{10}{12}\selectfont ignacio.poncearancibia@ucdconnect.ie}
    \and
    
    \IEEEauthorblockN{{\fontsize{12}{14}\selectfont Daniel Dotta}}
    \IEEEauthorblockA{
    \textit{\fontsize{10}{12}\selectfont Universidade Estadual de Campinas}\\
    \fontsize{10}{12}\selectfont Campinas, Brazil \\
    \fontsize{10}{12}\selectfont dottad@unicamp.br}
    \and

    \IEEEauthorblockN{{\fontsize{12}{14}\selectfont Federico Milano}}
    \hspace{18.5cm} \IEEEauthorblockA{
    \textit{\fontsize{10}{12}\selectfont University College Dublin}\\
    \fontsize{10}{12}\selectfont Dublin, Ireland \\
    \fontsize{10}{12}\selectfont federico.milano@ucd.ie}
}
\maketitle
\thispagestyle{fancy}

\begin{abstract}
  This paper introduces a novel formulation to evaluate the local synchronization of power system devices, namely \textit{Synchronization Energy} (SE).  The formulation is derived based on the complex frequency concept and the Teager Energy Operator applied to the complex power. This formulation offers valuable insights into the relationship between complex frequency of voltage and current of the device and its stationary operating. Based on this relationship we derive the conditions for a novel definition of local synchronization of power system devices. Through various case studies, the paper demonstrates how SE can effectively assess local synchronization under diverse operating conditions.
\end{abstract}


\section{Introduction}

\subsection{Motivation}

As power system dynamics evolve and become more complex due to the commissioning of more and more inverter-interfaced resources, it becomes essential to account for the interactions between different generating resources and their controls.  The synchronization mechanism is recognized to be an important aspect of these interactions and, in recent years, the definition and evaluation of device synchronization in the context inverter-dominated grids has been object of significant research interest~\cite{Junru2023, Peng2024, ponce2024localsync, He2024}.  In this framework, here we introduce a novel metric based on the Teager Energy Operator (TEO)~\cite{Kaiser_01} and the complex frequency approach~\cite{CF_Milano}, aimed at evaluating the local synchronization of power system devices.

\subsection{Literature Review}

In conventional transient stability analysis, system-wide stability and the loss of synchronism of a machine or a cluster of machines are considered synonyms \cite{Pavella:2000, Chiang:2010}.  More recently, the Kuramoto model, originally developed to study the dynamics of coupled oscillators, has been widely used to analyze frequency synchronization in power systems~\cite{florian2014, motter2013, Guo2021}.  However, this approach typically relies on reduced models that neglect voltage amplitude dynamics, limiting its applicability to real-world power systems. To address this, an \textit{augmented synchronization} framework was proposed in~\cite{Peng2024}, which incorporates both voltage amplitude and frequency dynamics into the analysis. Although this framework offers a more comprehensive view of system behavior, it is primarily tied to transient stability analysis, restricting the insights derived in the framework.

In the broader context of complex networks, the concept of \textit{network synchronizability} has been extensively studied~\cite{Jinhu2004, Liu2014, Jafarizadeh2021}.  Synchronizability quantifies a network's ability to maintain robust synchronization under varying conditions. This is often evaluated using an index based on the spectral analysis of the network Laplacian matrix. In~\cite{yu2024}, this approach was applied to assess synchronization within microgrids, focused on the frequency synchronization problem.  However, these studies focus primarily on global synchronizability and do not consider local interactions between power system devices and the grid.  In this work, we aim to bridge this gap by evaluating local synchronization, specifically addressing the local synchronizability from the perspective of complex power dynamics.

TEO has been applied in power systems~\cite{Kaiser2012, Nwobu2017, Sahoo2021, Reza2021, Zamora2023}, primarily for detecting low-frequency oscillations and estimating oscillation frequencies.  For example, in~\cite{Nwobu2017} and~\cite{Sahoo2021}, the energy operator is applied to voltage measurements to estimate grid frequency and improve device synchronization under varying conditions.  Similarly, in~\cite{Zamora2023}, the TEO is used to estimate oscillation frequencies during disturbances based on voltage signals.  Despite these applications, the use of TEO within the power systems community is confined to voltage measurements.  To the best of the authors' knowledge, no previous work has explored its application to other quantities, such as power.  In this work, we extend the application of TEO to complex power in power system devices.  Building on the complex frequency concept introduced in~\cite{CF_Milano}, we derive an analytical equation that offers a novel interpretation of the TEO of the complex power, allowing for a more generalized and insightful approach to evaluating local synchronization.

\subsection{Contributions}

The contributions of this paper are as follows.

\begin{itemize} 

\item A novel expression is introduced, based on the complex frequency of voltage and current, to quantify the TEO of complex power, hereinafter referred to as \textit{Synchronization Energy} (SE).

\item Using the proposed SE, we explicitly derive the general conditions for local synchronization of power system devices, which depend on the frequency and magnitude of voltage and current at the device's connection bus.

\item  Based on the proposed conditions of local synchronization, we propose a discussion on the difference between local synchronization and system-wide stability.

\end{itemize}

The information provided with the proposed SE definition is validated through a comprehensive case study based various test networks and several scenarios.  Results demonstrate the sensitivity of SE to device and grid parameters such as inertia and damping and reactance, as well as to the system operating conditions.

\subsection{Organization}

The remainder of the paper is organized as follows.  Section~\ref{sec:CF} introduces the complex frequency operator and its application to voltage and current of a power system. Section~\ref{sec:TEO} reviews fundamental of TEO and its interpretation. Section~\ref{sec:TEO_CP} introduces the concept of SE and its relationship with local synchronization of power system devices.  Numerical simulation results are discussed in Section~\ref{sec:results}.  Finally, conclusions are drawn in Section~\ref{sec:conclusion}.

\section{Background}
\label{sec:background}

In this section, the complex frequency (CF) concept and its application in power system analysis proposed in~\cite{CF_Milano} is introduced, as well as the Teager Energy Operator (TEO).

\subsection{Complex Frequency}
\label{sec:CF}

Consider a complex signal $\bar x(t) = a(t)e^{\jmath\phi(t)}$, where $a(t)$ is the time varying amplitude and $\phi(t)$ the signal phase. The time derivative of $\bar x(t)$ is given as follows:
\begin{equation}
    \label{eq:dot_x}
    \dot{\bar{x}} = \left( \frac{\dot a}{a} + \jmath\dot \phi \right)\bar x,
\end{equation}
where time dependency has been omitted for simplicity of notation. Defining $u=\ln(a)$, we can rewrite~\eqref{eq:dot_x} as follows:
\begin{align}
    \label{eq:def_cf}
    \dot{\bar{x}} &= \left( \dot u + \jmath\dot \phi \right)\bar x = \bar \eta\bar x,
\end{align}
where $\bar\eta$ is the \textit{complex frequency} (CF) of $\bar{x}$. The real part is the \textit{instantaneous bandwidth} and the imaginary part is the \textit{instantaneous frequency}. Hence, the CF can be interpreted as the generalization of the instantaneous frequency for a signal with time varying amplitude.

\subsubsection{Complex Frequency in Power Systems}

A relevant application of the complex frequency concept are power systems quantities such as voltages and currents \cite{CF_Milano}.  Let us consider the time-dependent Park vector of voltage $\bar{v}(t)$ and injected current $\bar{\imath}(t)$ at a particular bus as a complex value in the dq-axis reference. These quantities can be expressed in their polar coordinates, as follows:
\begin{align}
    \label{eq:cf_v}
    \bar{v} &= v_d + \jmath v_q = v(t)e^{\jmath \theta_v(t)} \\
    \label{eq:cf_i}
    \bar{\imath} &= \imath_d + \jmath\imath_q = \imath(t)e^{\jmath \theta_{\imath}(t)}.
\end{align}

Differentiating~\eqref{eq:cf_v} and~\eqref{eq:cf_i} as in~\eqref{eq:def_cf}, one obtains:
\begin{align}
    \label{eq:cf_v_2}
    \dot{\bar{v}} &= (\rho_v + \jmath \omega_v)\bar{v} = \bar{\eta}_v\bar{v} \\
    \label{eq:cf_i_2}
     \dot{\bar{\imath}} &= (\rho_{\imath} + \jmath \omega_\imath)\bar{\imath} = \bar{\eta}_\imath\bar{\imath}, 
\end{align}
where $\bar{\eta}_v$ and $\bar{\eta}_\imath$ are the complex frequency of voltage and current, respectively. The instantaneous frequency and instantaneous bandwidth of the CF of voltage and current can be written as:
\begin{align}
    \rho_v &= \frac{\dot{v}}{v}, \quad \omega_v = \dot{\theta}_v \\
    \rho_{\imath} &= \frac{\dot{\imath}}{\imath}, \quad \omega_\imath = \dot{\theta}_{\imath}.
\end{align}

Note that $\bar{\eta}_v$ and $\bar{\eta}_\imath$ are frequency variations, that is, in ideal conditions $\rho_v \xrightarrow{} 0$, $\rho_{\imath} \xrightarrow{} 0$, $\omega_v \xrightarrow{} \omega_n$, and $\omega_\imath \xrightarrow{} \omega_n$, where $\omega_n = 2\pi f_n$ is the nominal system frequency. Also, in general, during a transient condition, $\bar{\eta}_v \neq \bar{\eta}_\imath$. The exception is when the device is a constant impedance load, where $\bar{\eta}_v = \bar{\eta}_\imath$~\cite{CF_Milano}.

\subsection{Teager Energy Operator}
\label{sec:TEO}

TEO, also known as Teager-Kaiser Energy Operator, is an non-linear operator designed to track the energy of the source of a signal, specifically, the energy required to generate the signal~\cite{Kaiser_01}. In continuous time, the TEO is defined by
\begin{equation}
    \label{eq:TEO_continuo}
    \psi(x(t)) = \dot{x}^2(t) - x(t) \, \ddot{x}(t),
\end{equation}
where $x(t)$ is the measured signal, and $\dot{x}(t)$ and $\ddot{x}(t)$ are the first and second derivatives of $x(t)$, respectively. If the measured signal $x(t)$ is a position, the TEO can be directly related to the total energy per unit of mass, that is, the sum of kinetic and potential energy.  

An additional interpretation of the TEO can be obtained by the Lie bracket operator. The Lie bracket measure the instantaneous difference in the relative rate of change between two signals $x$ and $y$, defined as follows~\cite{Maragos1995}:
\begin{equation}
    \label{eq:lie}
    [x,y] = \dot{x}y - x\dot{y}
\end{equation}

\noindent where [*] is the Lie bracket operator. For $y=\dot{x}$, the Lie bracket is the continuous TEO defined in~\eqref{eq:TEO_continuo}. Therefore, for any signal, the TEO can be interpreted as the instantaneous difference in the relative rate of change of $x(t)$ and $\dot{x}(t)$~\cite{Maragos1995}.

The continuous-time TEO~\eqref{eq:TEO_continuo} can be extended to complex signals~\cite{Hamila_ambiguity}, as follows:
\begin{equation}
    \label{eq:complex_TEO}
    \psi_c(\bar{x})= \dot{\bar{x}}^*\dot{\bar{x}} - 
    \frac{1}{2}\left( \ddot{\bar{x}} \, \bar{x}^* + \bar{x} \, \ddot{\bar{x}}^* \right),
\end{equation}
where $\bar{x} = x_r + jx_i$.  Note that the TEO of a complex signal can be written as the the sum of the continuous-time TEO of the real and imaginary parts:
\begin{equation}
    \label{eq:complex_TEO2}
    \psi_c(\bar{x}) = \psi(x_r) + \psi(x_i).
\end{equation}

The TEO does not measure the energy of the measured signal over a time interval, but its \textit{instantaneous energy} required to generate the signal.  For instance, for a exponentially-decaying signal $e^{-\alpha t}$, the TEO calculated by~\eqref{eq:TEO_continuo} is zero.  Another important aspect of the TEO is that it takes into account both the amplitude and frequency of the signal to define its instantaneous energy.

\section{Synchronization Energy}
\label{sec:TEO_CP}

In this section, we derive the complex frequency-based TEO of the complex power of a device, and discuss its application in the evaluation of local synchronization of power system devices.

\subsection{Teager Energy Operator of Complex Power}

The instantaneous complex power injected by a device at a bus is calculated using Park vectors as follows
\begin{align}
    \label{eq:sh}
    \begin{aligned}
    \bar{s} &= p + jq \\ 
    &= \bar{v} \, \bar{\imath}^*,
    \end{aligned}
\end{align}
where $\bar{v}$ and $\bar{\imath}^*$ are the voltage of the bus and conjugated current injected in the bus, respectively. The derivative of~\eqref{eq:sh} can be calculated using the CF concept, as follows
\begin{align}
    \label{eq:derivative_sh}
    \begin{aligned}
    \dot{\bar{s}} &= \bar{\eta}_v\bar{s} + \bar{\eta}_i^*\bar{s} \\
    & = (\bar{\eta}_v + \bar{\eta}_i^*)\bar{s},
    \end{aligned}
\end{align}

\noindent where $(\bar{\eta}_v + \bar{\eta}_i^*)$ is the CF of the complex power, and $\bar{\eta}_v$ and $\bar{\eta}_i$ are the ones defined in~\eqref{eq:cf_v_2} and~\eqref{eq:cf_i_2}, respectively.  To calculate the TEO of the complex power, the second time derivative of~\eqref{eq:sh} is needed, which can be expressed in terms of the square of its CF and the derivative of the CF:
\begin{align}
    \label{eq:2dot_sh}
    \begin{aligned}
    \ddot{\bar{s}} &= (\dot{\bar{\eta}}_v + \dot{\bar{\eta}}_i^*)\bar{s} + (\bar{\eta}_v + \bar{\eta}_i^*)\dot{\bar{s}} \\
    &= (\dot{\bar{\eta}}_v + \dot{\bar{\eta}}_i^*)\bar{s} + (\bar{\eta}_v + \bar{\eta}_i^*)(\bar{\eta}_v + \bar{\eta}_i^*)\bar{s} \\
    &= [(\dot{\bar{\eta}}_v + \dot{\bar{\eta}}_i^*) + (\bar{\eta}_v + \bar{\eta}_i^*)^2]\bar{s}.
    \end{aligned}
\end{align}

Then, the TEO of the injected complex power can be calculated using \eqref{eq:complex_TEO} and substituting \eqref{eq:derivative_sh} and \eqref{eq:2dot_sh} into it, as follows
\begin{align}
    \label{eq:TEO_sh}
    \begin{aligned}
    \psi_c(\bar{s}) &= \dot{\bar{s}}^*\dot{\bar{s}} - \frac{1}{2}[\ddot{\bar{s}} \, \bar{s}^* + \bar{s} \, \ddot{\bar{s}}^*] \\
     &= (|\bar{\eta}_v|^2 + |\bar{\eta}_\imath|^2 + 2\rho_v\rho_\imath -2\omega_v\omega_\imath)|\bar{s}|^2 \\
     &- \frac{1}{2}\left[ 2\Re\{ \dot{\bar{\eta}}_v \} + 2\Re\{ \dot{\bar{\eta}}_\imath \} +  2\Re\{ \bar{\eta}_v + \bar{\eta}_\imath^* \} \right]|\bar{s}|^2 \\
    &= \left( 2\left( \omega_v - \omega_\imath\right)^2 -\Re\{ \dot{\bar{\eta}}_v \} - \Re\{ \dot{\bar{\eta}}_\imath\}   \right)|\bar{s}|^2 ,
    \end{aligned}
\end{align}
\noindent where
\begin{align}
    \label{eq:|s|}
    |\bar{s}| &= \sqrt{p^2 + q^2} , \\
    \label{eq:dot_etav}
    \Re\{ \dot{\bar{\eta}}_v\} &= \left( \frac{\dot{v}}{v}  \right)' = \frac{\ddot{v}}{v} - \rho_v^2,\\
    \label{eq:dot_etai}
    \Re\{ \dot{\bar{\eta}}_\imath\} &= \left( \frac{\dot{\imath}}{\imath}  \right)' = \frac{\ddot{\imath}}{\imath} - \rho_\imath^2.
\end{align}

Substituting~\eqref{eq:dot_etav} and~\eqref{eq:dot_etai} into~\eqref{eq:TEO_sh}, one obtains 
\begin{equation}
    \label{eq:final_TEO_1}
      \psi_c(\bar{s}) = \left( 2\left( \omega_v - \omega_\imath\right)^2 + \rho_v^2 - \frac{\ddot{v}}{v} + \rho_\imath^2  - \frac{\ddot{\imath}}{\imath} \right)|\bar{s}|^2.
\end{equation}

The terms that are dependent of the real part of the CF of voltage and current and the respective second time derivative can be rewritten using the definition of TEO applied to a real signal~\eqref{eq:TEO_continuo}, as follows:
\begin{equation}
    \label{eq:final_TEO_2}
       \psi_c(\bar{s}) = \left[ 2\left( \omega_v - \omega_\imath\right)^2  +  \frac{\psi(v)}{v^2}   + \frac{\psi(\imath)}{\imath^2} \right]|\bar{s}|^2,
\end{equation}
\noindent where $\psi(v)$ and $\psi(\imath)$ are the TEO of the magnitude of the voltage and current, respectively. 

It is convenient to separate~\eqref{eq:final_TEO_2} in two components that resemble a joint time–frequency distribution (TFD) analysis of the signal. Considering the Wigner TFD (see the Appendix), where the first conditional moment in frequency is the instantaneous frequency and given the definition of the conditional standard deviation of the distribution in~\eqref{eq:conditional_3}, one can rewrite \eqref{eq:final_TEO_2} as follows:
\begin{equation}
    \label{eq:final_TEO}
      \fbox{ $\psi_c(\bar{s}) = \left( \omega_v - \omega_\imath\right)^2 2|\bar{s}|^2 + \left( \sigma_{\omega_v | t}^2   + \sigma_{\omega_{\imath} | t}^2   \right) 2 |\bar{s}|^2 $ }
\end{equation}

\noindent where $\sigma_{\omega_v | t}^2$ and $\sigma_{\omega_\imath | t}^2$ are the conditional standard deviation in frequency of the magnitude of voltage and current, respectively, expressed as follows:
\begin{equation}
    \label{eq:sigma_v}
    \sigma_{\omega_v | t}^2 = \frac{1}{2}\left[ 
 \left( \frac{\dot{v}}{v} \right)^2 - \frac{\ddot{v}}{v} \right] , \qquad  \sigma_{\omega_\imath | t}^2 = \frac{1}{2}\left[ 
 \left( \frac{\dot{\imath}}{\imath} \right)^2 - \frac{\ddot{\imath}}{\imath} \right].
\end{equation}

The first term in \eqref{eq:final_TEO} is based on the first conditional moment in frequency of Wigner TFD, and the second term is based on the second conditional moment in frequency.  Normalizing~\eqref{eq:final_TEO} by $2|\bar{s}|^2$, and subtracting the square of instantaneous frequency of $\bar{s}(t)$, the conditional standard deviation of $\bar{s}(t)$ can be written as:
\begin{equation}
    \label{eq:sigma2_sh}
    \sigma^2_{\omega_s|t}  =   \sigma_{\omega_v | t}^2   + \sigma_{\omega_\imath | t}^2 \, .
\end{equation}

Equation~\eqref{eq:final_TEO} is the sought expression that directly links the TEO of the complex power of a particular device with the instantaneous frequency of the complex power and its conditional standard deviation in frequency.  The SE is defined based on \eqref{eq:sigma2_sh}, as follows.

\vspace{0.1cm}

\noindent \textbf{Definition 1.} \textit{(Synchronization Energy): The energy related to the oscillations in active and reactive power of a device $h$, namely synchronization energy, is obtained by the TEO of the complex power $\psi_c(\bar{s})$ formulated in~\eqref{eq:final_TEO}.}

\vspace{0.1cm}

The following three remarks are relevant for the interpretation of the results given in the next section.

\textit{Remark 1.1:} $\psi_c(\bar{s})$ is not directly linked to the kinetic and potential energy of the device.  However, $\psi_c(\bar{s})$ measures the instantaneous difference in the relative rate of change between $\bar{s}$ and $\bar{\dot{s}}$.  Therefore, a large value of $\psi_c(\bar{s})$ indicates that the complex power is experiencing significant acceleration/deceleration; a small change indicates a smooth variation; and zero indicates steady state.

\textit{Remark 1.2:} The variations in active and reactive power arise due to the discrepancy between $\omega_v$ and $\omega_\imath$, as well as the rate of change in the magnitude of the device's voltage and current.  

\textit{Remark 1.3:} Since the Teager Energy of a signal represents the energy required to generate it, the proposed synchronization energy quantifies the energy a device requires to maintain synchronism (or to attempt to do so). A high value of this metric indicates that the device exerts greater effort to stay synchronized or to regain synchronization.

\subsection{Local Synchronization}

The proposed SE can be viewed as a metric to evaluate the variations in the complex power of a particular device, encompassing the active and reactive power.  Based on this observation and in the formulation based on the complex frequency in~\eqref{eq:final_TEO} we can explicitly obtain the conditions for local synchronization of the device, based solely on quantities measured in the point of connection with the system.

\noindent \textbf{Definition 2.} \textit{(Local Complex Power Synchronization): Consider a power system device injecting power in the node $h$.  The device achieve complex power synchronization if its synchronization energy asymptotically converge to zero, i.e., $\psi_c(\bar{s}) \xrightarrow{} 0$, $t \xrightarrow{} \infty$.}

\textit{Remark 2.1}: The condition $\psi_c(\bar{s}) = 0$  occurs when the complex power of the device is not oscillating.  Based on~\eqref{eq:final_TEO}, this occurs for $\omega_v = \omega_\imath$ and $\sigma^2_{\omega_s|t} = 0$.  The first one is called isofrequential condition and the second one the stationary condition.  The stationary condition is mostly related to constant voltage and current, i.e., $\rho_v = 0$ and $\rho_{\imath} = 0$. 

These conditions are same as those proposed in~\cite{ponce2024localsync} for the local synchronization of a device.  However, the definition based on TEO requires only a single condition to be satisfied, namely $\psi_c(\bar{s}) \xrightarrow{} 0$, whereas, in \cite{ponce2024localsync}, the conditions are three.  Moreover, the stationary operating condition $\dot{v} \xrightarrow{} 0$ is referred to as \textit{augmented synchronization} in~\cite{Peng2024}. However, in \eqref{eq:final_TEO}, the isofrequential and stationary conditions are explicitly presented. Additionally, unlike the CF of the dynamic equivalent admittance proposed in~\cite{ponce2024localsync}, SE considers the sum of variations in the voltage and current magnitudes in the conditional standard deviation of the complex power~\eqref{eq:sigma2_sh}.

\textit{Remark 2.2}: The condition $\sigma^2_{\omega_s|t} = 0$ can also occur during a steady-state oscillation in the voltage and current magnitudes, if $\sigma_{\omega_v | t}^2 = -\sigma_{\omega_\imath | t}^2 $.

\section{Case studies}
\label{sec:results}

This section discusses the use of SE to evaluate the local synchronization in power systems. The following benchmark systems are considered: (i) single machine infinite bus, (ii) Kundur two-areas system, (iii) IEEE 14-bus system, and (iv) GFL inverter-based system.  Simulations are obtained using Dome, a Python-based power system software tool~\cite{Milano_python}. 

The test cases presented in Sections \ref{sub:omib} and \ref{sub:kundur} are utilized to distinguish between local synchronization and stability and to demonstrate how the proposed SE can be applied and interpreted in these different scenarios.  The test case of Section \ref{sub:14bus} serves two purposes: validate the proposed CF formulation by comparing it with numerical estimation approach, and second, to illustrate how the SE metric can be used to assess local robustness. Finally, the test case of Section \ref{sub:ibr} demonstrates the application of the SE in an IBR system.

\begin{figure}[htb]
    \includegraphics[width=0.99\linewidth]{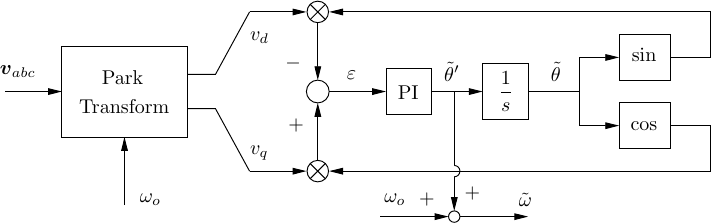}
    \centering
    \caption{Scheme of the synchronous reference frame PLL.}
    \label{fig:srfpll}
\end{figure}

The estimations of CF variables and the TEO of voltage and current magnitude are obtained using synchronous-reference frame Phase Locked Loop (SRF-PLL) and voltage and current measurements at the point of connection of the device under analysis.  A fundamental model of the SRF-PLL is presented in Figure \ref{fig:srfpll}, which consists of a phase detector (implemented as the Park transform), a loop filter (implemented as a PI control), and a voltage-controlled oscillator (implemented as an integrator).  The output of the PI control, namely $\tilde{\theta}'$, is an estimation of the frequency deviation with respect to the synchronous reference frame $\omega_o$.  In the simulations, the proportional and integral gains of the PI controller are set to 10 and 20, respectively.

\subsection{Single-Machine Infinite-Bus (SMIB) system}
\label{sub:omib}

\begin{figure}[htb]
    \includegraphics[width=0.80\linewidth]{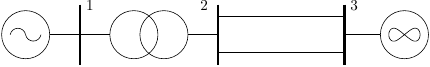}
    \centering
    \caption{Single line diagram of a SMIB system.}
    \label{fig:SMIB}
\end{figure}

The system is composed of a synchronous machine represented with a 2nd order model connected to an infinite bus, as shown in Figure~\ref{fig:SMIB}.  This system is used to evaluate the sensitivity of the SE of the machine regarding the parameters of: inertia, damping, and reactance.  For the following cases, the disturbance consists of a short circuit applied at bus 3 at $t=1$ s, and cleared at $t=1.1$ s.

\subsubsection{Inertia sensitivity}

\begin{figure}[htb]
    \includegraphics[width=0.9\linewidth]{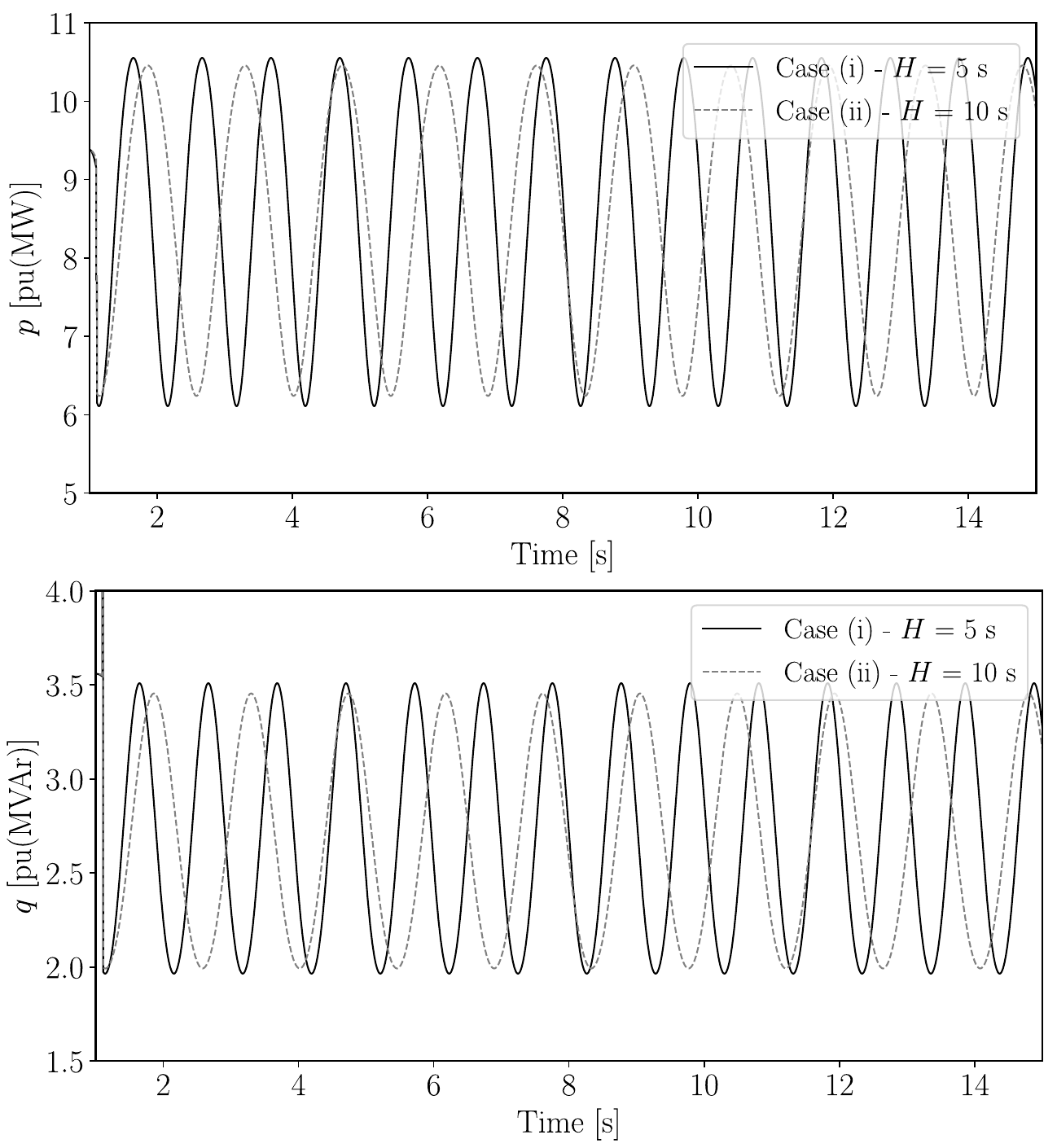}
    \centering
    \caption{SMIB system --- Active and reactive powers injected at bus 1 for two values of inertia.}
    \label{fig:SMIB_p_h}
\end{figure}

We first analyze the system without damping ($D = 0$) for two inertia cases: (i) $H = 5$ s and (ii) $H = 10$ s.  The active and reactive power responses for both cases are shown in Figure~\ref{fig:SMIB_p_h}.  Although the system remains stable, it exhibits undamped oscillations.  Note that the impact of inertia is not directly observable in the magnitude of the active and reactive power. On the other hand, we observe the impact on frequency oscillation.

\begin{figure}[htb]
    \includegraphics[width=0.9\linewidth]{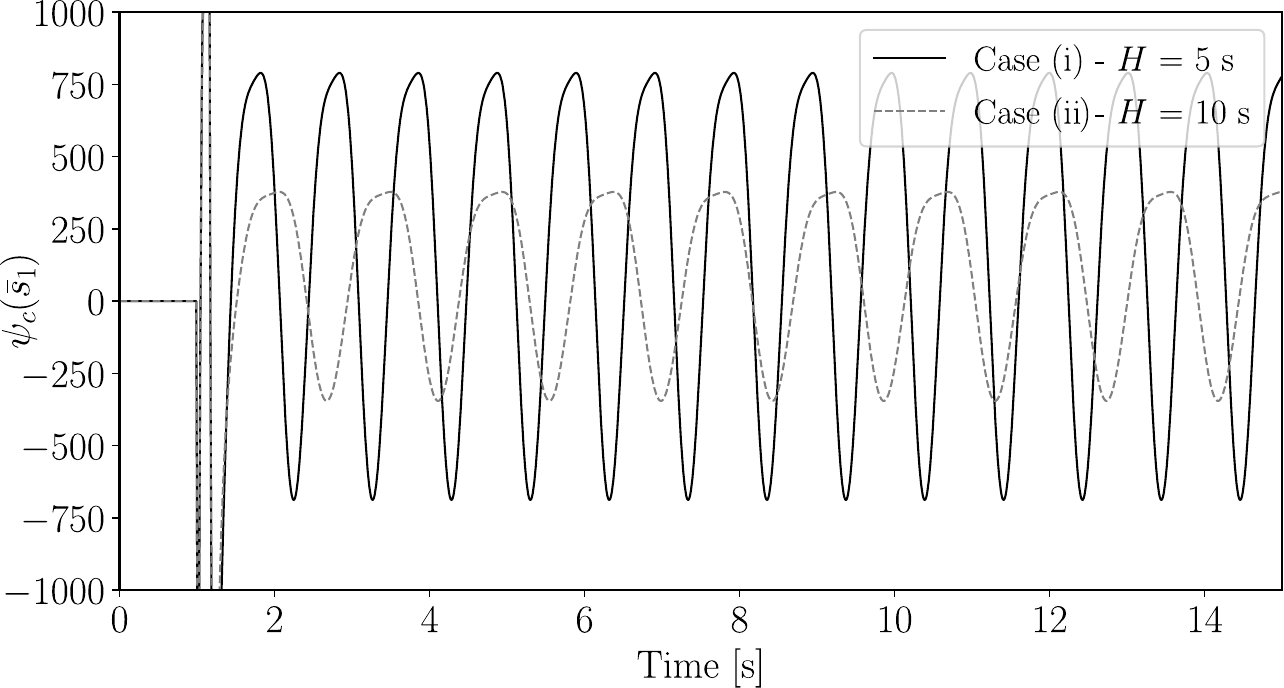}
    \centering
    \caption{SMIB system --- Comparison of SEs at bus 1 for two values of inertia. }
    \label{fig:SMIB_syncenergy_h}
\end{figure}

Figure~\ref{fig:SMIB_syncenergy_h} shows the SE of the machine for both cases.  In neither case does the machine reach asymptotic local synchronization, as $\psi_c(\bar{s}_1) \neq 0$.  The inertia has a direct impact on the SE, with high-inertia case (ii) showing lower SE compared to low-inertia case (i).  Based on this, we can conclude that although neither machine reaches asymptotic local synchronization, the machine in case (ii) requires less energy to attempt to maintain local synchronism.


\subsubsection{Damping} 

\begin{figure}[htb!]
    \includegraphics[width=0.9\linewidth]{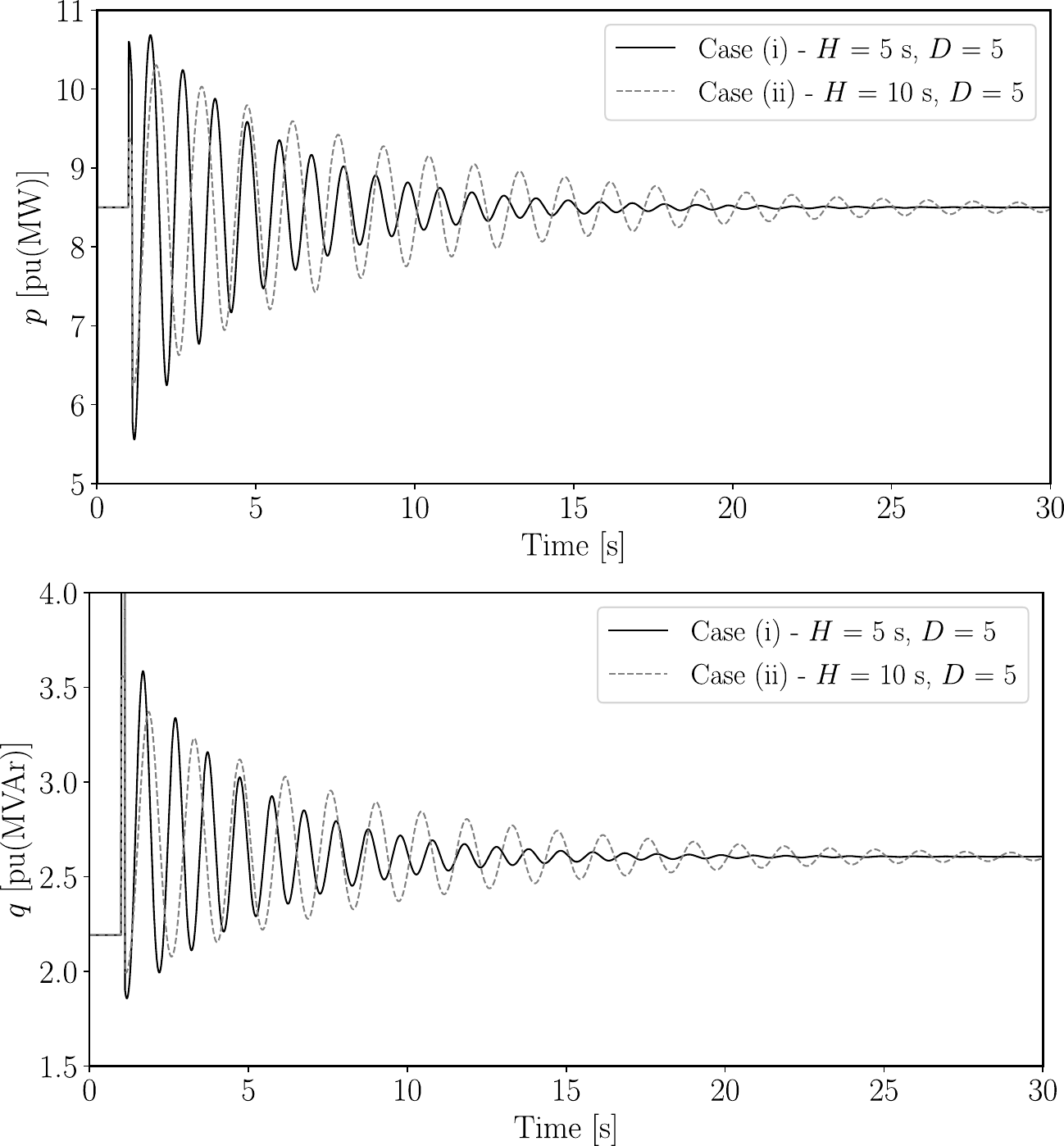}
    \centering
    \caption{SMIB system --- Active and reactive powers injected at bus 1 for two values of inertia and $D = 5$.}
    \label{fig:SMIB_p_d}
\end{figure}

We consider again the impact of inertia, but including a damping factor of $D = 5$.  The active and reactive power responses of the machine are shown in Figure \ref{fig:SMIB_p_d}.  Notably, the complex power of the machine in case (i) exhibits greater initial variations following the disturbance but reaches a steady-state operating condition more quickly. This is expected, as a generator with lower inertia has less stored kinetic energy to dissipate.  For the SMIB system, for example, the impact of inertia variation in the damping and frequency of the oscillation mode can be easily obtained by observing the eigenvalue of the system: $\lambda = -D/2H \pm \sqrt{K_s\Omega/2H}$.  On the other hand, case (ii), with higher inertia, experiences more prolonged oscillations and takes longer to reach steady state. From the perspective of the TEO, both the magnitude and frequency of oscillations are considered when calculating energy.

\begin{figure}[htb!]
    \includegraphics[width=0.9\linewidth]{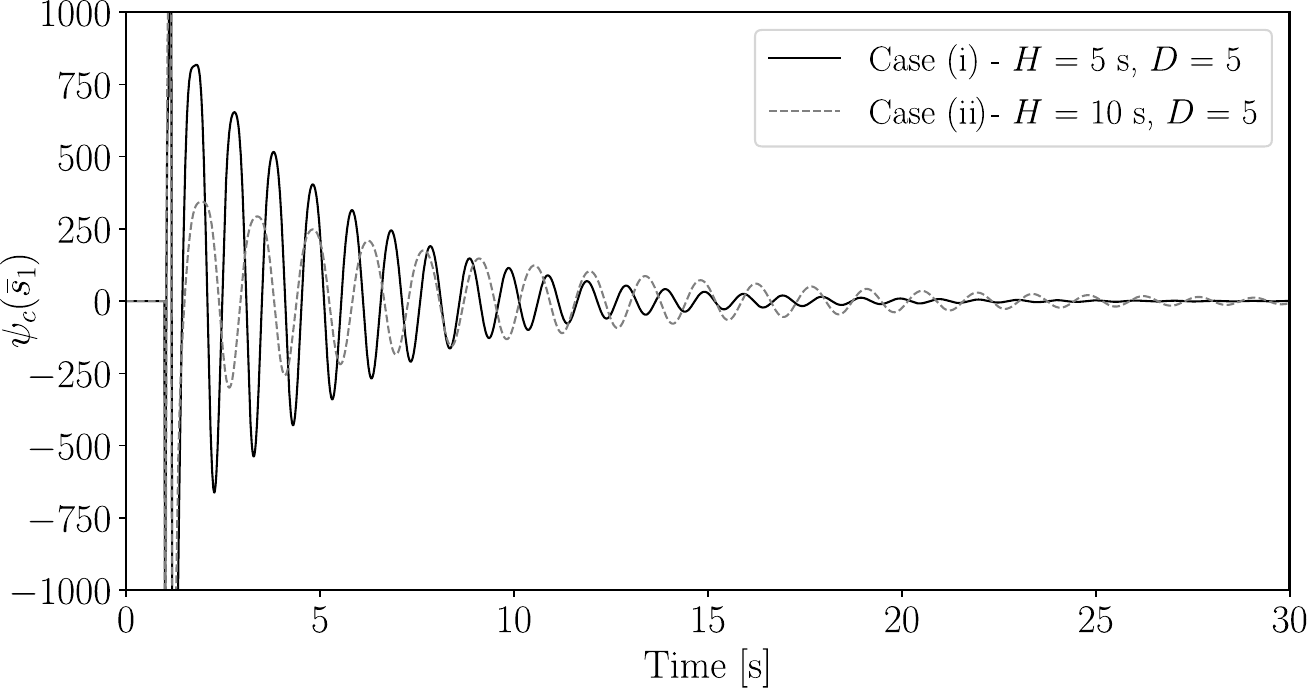}
    \centering
    \caption{SMIB system --- Comparison of SEs at bus 1 for two values of inertia and $D = 5$.}
    \label{fig:SMIB_syncenergy_d}
\end{figure}

Figure~\ref{fig:SMIB_syncenergy_d} shows the SE for both cases. During the transient period following the disturbance, case (ii) exhibits the lowest SE, which aligns with its higher inertia.  Since the oscillation frequency is lower in case (ii), the resulting SE is ultimately lower compared to case (i). However, the asymptotic local synchronization is achieved later in case (ii).  Therefore, we can conclude that while the machine in case (ii), with higher inertia, requires less energy to maintain synchronism, it takes longer to achieve asymptotic local synchronization.

\subsubsection{Electrical distance}

\begin{figure}[htb!]
    \includegraphics[width=0.9\linewidth]{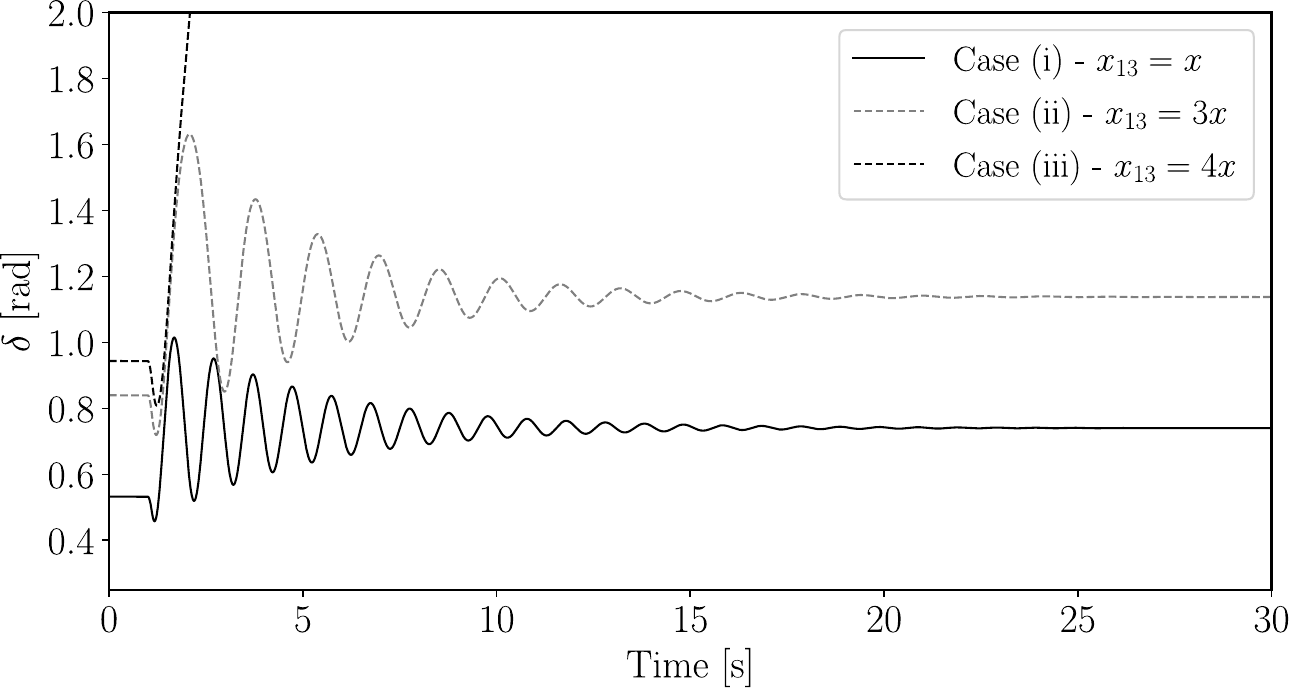}
    \centering
    \caption{SMIB system --- Machine rotor angle for three values of line reactance.}
    \label{fig:SMIB_ang_xt}
    %\vspace{-2mm}
\end{figure}

We now analyze three cases concerning the electrical distance between the machine and the infinite bus: (i) base case $x_{13} = x$, (ii) $x_{13} = 3x$, and (iii) $x_{13} = 4x$.  In all cases, a short circuit is applied at bus 1, with the machine having an inertia of $H = 5$ and damping of $D = 5$.  The machine rotor angles for the three cases are shown in Figure~\ref{fig:SMIB_ang_xt}, where angular instability in case (iii) is clearly visible after the first swing.

\begin{figure}[htb!]
    \includegraphics[width=0.9\linewidth]{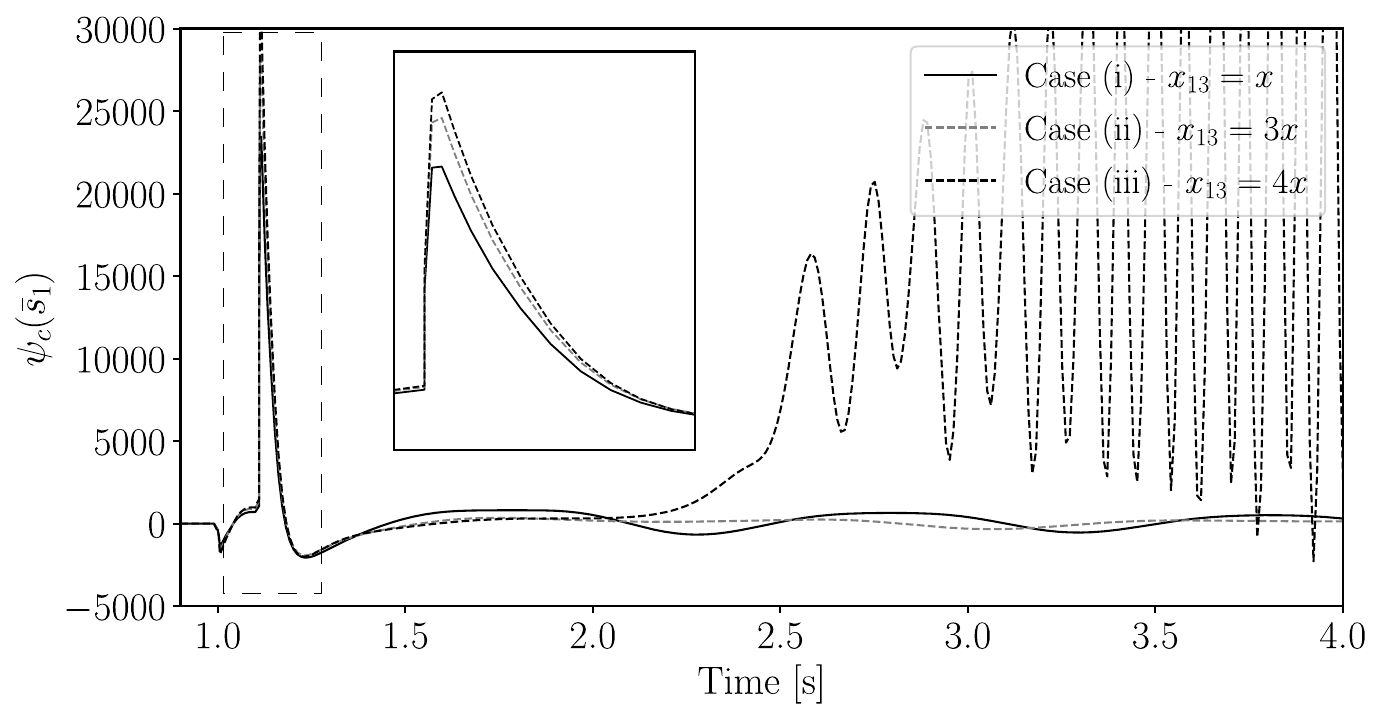}
    \centering
    \caption{SMIB system --- Comparison of SEs at bus 1 for the three values of line reactance.}
    \label{fig:SMIB_syncenergy_xt}
    %\vspace{-2mm}
\end{figure}

The SEs obtained for the various values of the line reactance are shown in Figure \ref{fig:SMIB_syncenergy_xt}.  In case (iii), the machine loses local synchronism, as $\psi_c(\bar{s}_1) \xrightarrow{} \infty$.  It is also evident that the highest synchronization energy following the short circuit occurs in case (iii), indicating that the energy required to maintain synchronism is highest in this weaker grid connection scenario.

It is important to note that the machine loses local synchronism after 2 s, as $\psi_c$ diverges, despite the angular instability observed immediately after the first swing.  This behavior  highlights the fact that angular stability and local synchronization are independent phenomena \cite{ponce2024localsync}.  

\begin{figure}[htb!]
    \includegraphics[width=\linewidth]{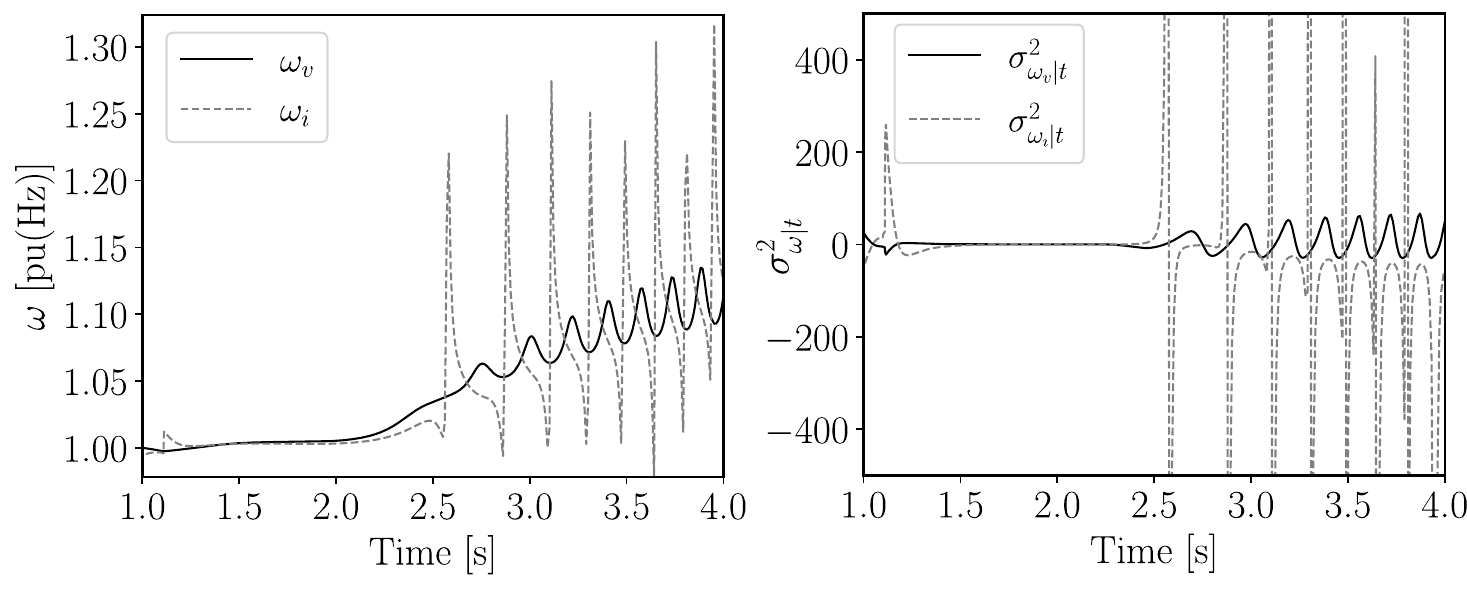}
    \centering
    \caption{SMIB system --- Components of SE at bus 1 for case (iii) of the electrical distance scenario.}
    \label{fig:SMIB_key_xt}
\end{figure}

The components of the SE, namely the imaginary part of the CF of voltage and current, along with the conditional standard deviation of the voltage and current magnitudes, are presented in Figure~\ref{fig:SMIB_key_xt}.  Notably, the conditions for local synchronism discussed in \textit{Remark 2.1} are not satisfied after 2 s. In this example, the loss of local synchronism results from variations in both the frequency and the magnitude of voltage and current.

\subsection{Kundur 2-areas system}
\label{sub:kundur}

We now use the two-area system to show the case where the system loses angular stability but the machines keep their local synchronism.  The system is composed by four machines modeled by the 4th-order model equipped with primary voltage regulators (AVRs).  The data of the system and the base-case operating condition can be found in~\cite{Kundur}.

\begin{figure}[htb!]
    \includegraphics[width=0.95\linewidth]{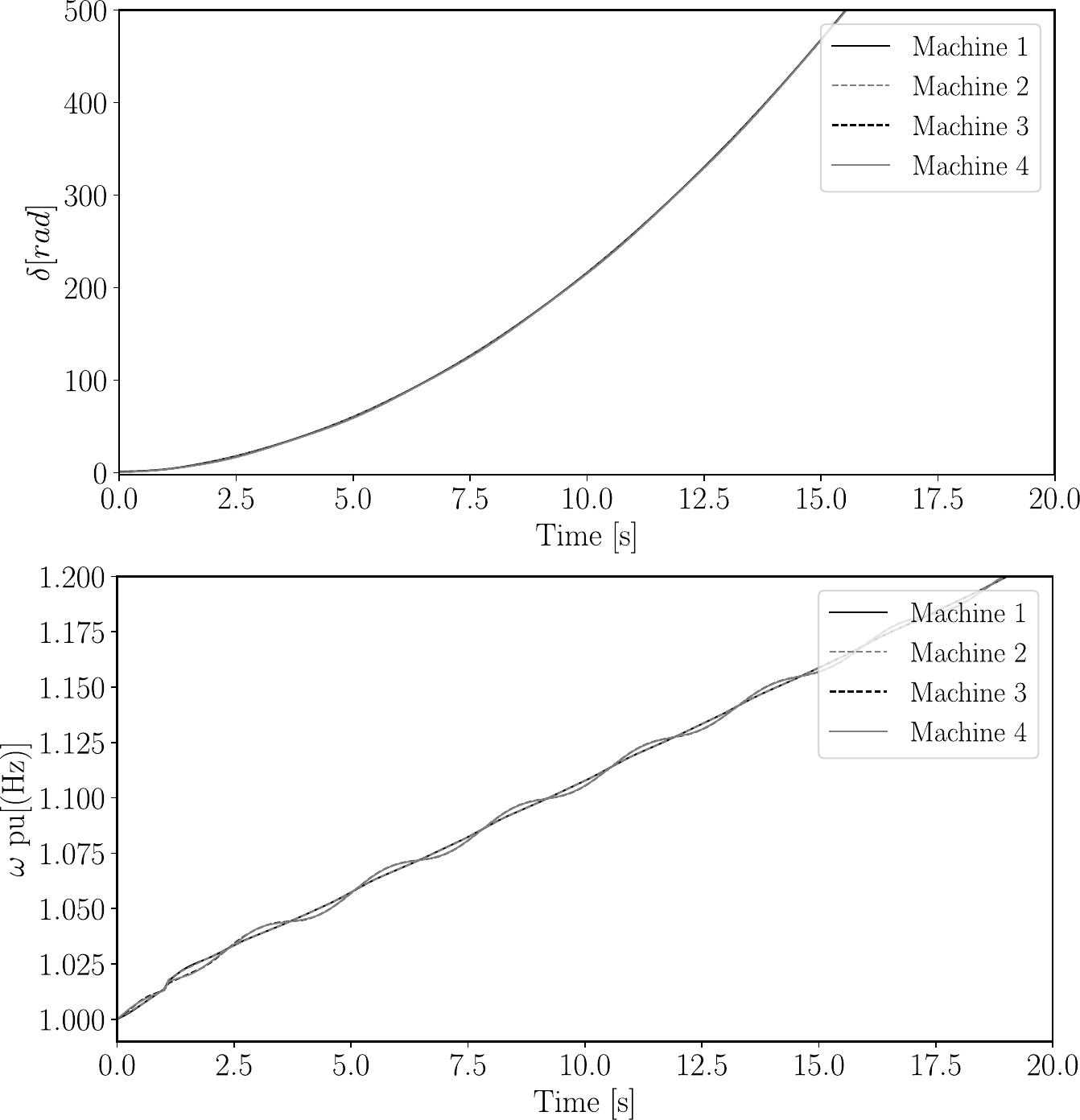}
    \centering
    \caption{Two-area system --- Machine rotor angles and internal frequency.}
    \label{fig:ang_kundur}
\end{figure}

The contingency consists of a three-phase short circuit applied in the middle of one of the circuits in the interconnection between the two areas.  The fault is applied at $t = 1$ s and cleared at $t = 1.12$ s by opening the faulted line.  The system is set up so that the machines do not have enough power reserve to compensate the power imbalances following the clearance of the fault.  As a result, this case illustrates a frequency instability, that is, the machines maintain local synchronism but are unstable as their rotor angle and speed show an unrecoverable drift (see Figure~\ref{fig:ang_kundur}).

\begin{figure}[htb!]
    \includegraphics[width=0.95\linewidth]{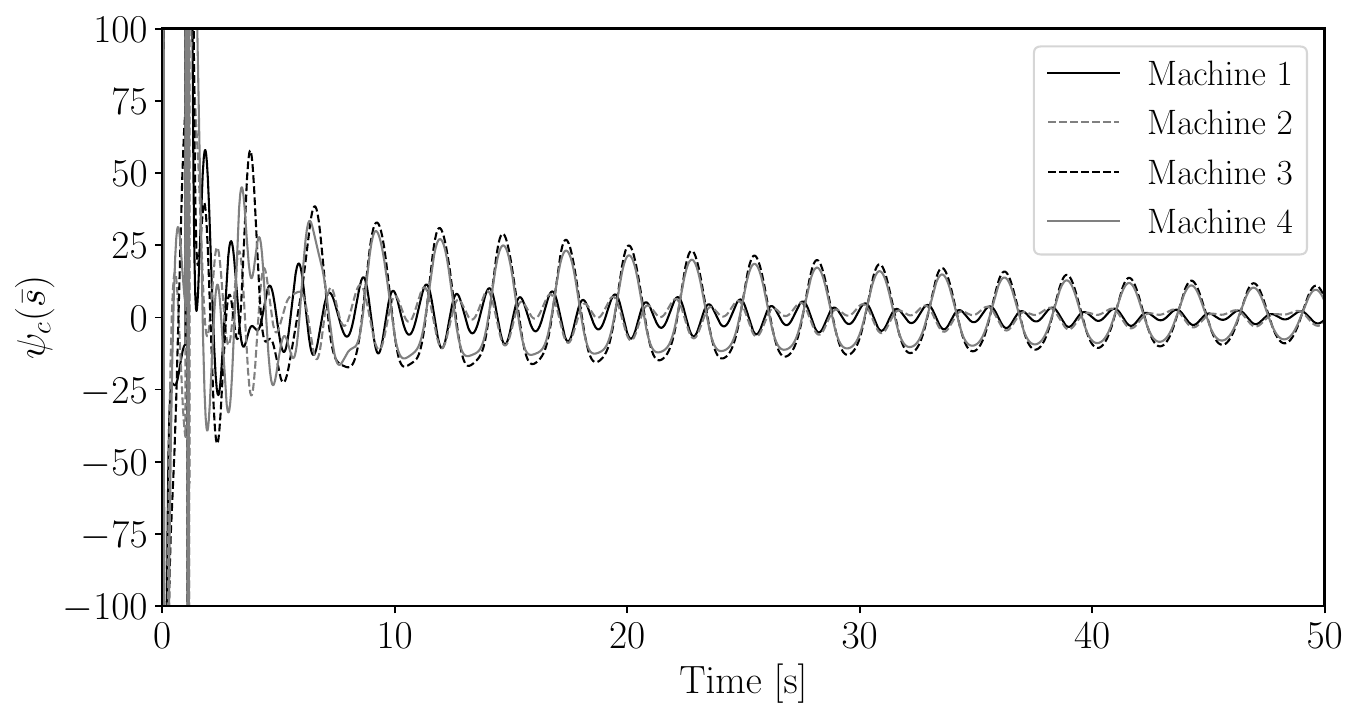}
    \centering
    \caption{Two-area system --- SE of machines.}
    \label{fig:syncenergy_kundur}
\end{figure}

The SE of the machines are shown in Figure~\ref{fig:syncenergy_kundur}.  Despite the system exhibiting instability, the SE remains bounded and converges to zero over a long simulation time. The oscillations in the SE are due to the presence of a lightly damped inter-area mode, since the system is without PSS.  Therefore, we can conclude that the machines achieve asymptotic local synchronism, as $\psi_c \xrightarrow{} 0$.  Additionally, the machine angles remain closely aligned, consistent with the fundamental definition of synchronism.  This example illustrates that system devices can be synchronized yet unstable.

\subsection{IEEE 14-bus system}
\label{sub:14bus}

The IEEE 14-bus system is a well-known system that consists of 2 synchronous machines and 3 synchronous compensators, represented by a 6th-order model. The system also includes primary voltage regulators (AVRs) and a PSS connected at machine 1.  All loads are modeled as constant power consumption.  The complete data of the system and the base operating point can be found in~\cite{MilanoBook1}. 

This system is used to evaluate the consistency of the proposed formulation of SE in \eqref{eq:final_TEO}, calculated based on the CF quantities. To evaluate the SE in different operating conditions, we consider the following cases: (i) stable steady-sate operating condition, and (ii) an operating condition with increased load consumption that leads to a stable limit cycle, that is, a steady-state oscillation with constant period.

\subsubsection{Stable operating point}

In this case, we consider the base operating point and the system is without PSS.  A three-phase short circuit is applied at bus 14 at simulation time $t = 1$ s and cleared at $t_{cl} = 1.12$  s. 

To show the consistency of our proposed formulation, the complex frequency-based formulation~\eqref{eq:final_TEO} is compared with the numerical estimation of the TEO of active and reactive power, obtained by using~\eqref{eq:complex_TEO2}, as follows:
\begin{equation}
    \label{eq:estimation_TEO}
    \psi_c(\bar{s}_1) = \psi(p_1) + \psi(q_1) = (\dot{p}_1^2 - p_1\ddot{p}_1) + (\dot{q}_1^2 - q_1\ddot{q}_1).
\end{equation}

\begin{figure}[htb!]
    \includegraphics[width=0.9\linewidth]{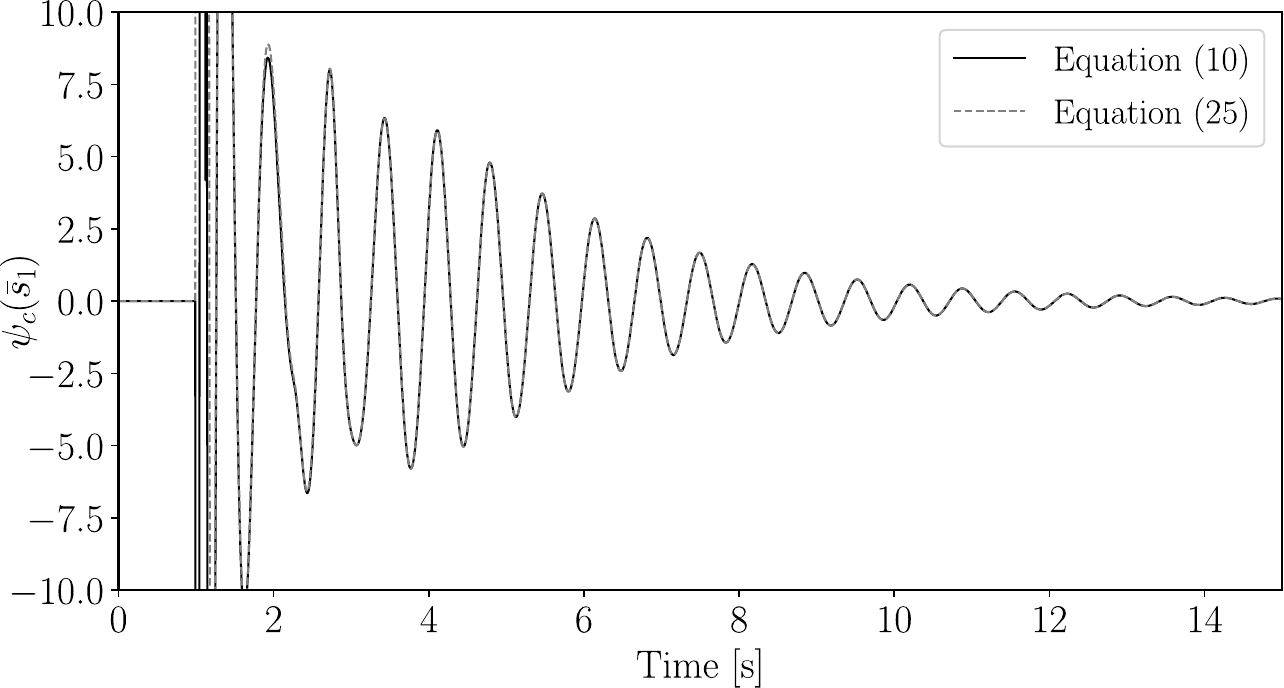}
    \centering
    \caption{IEEE 14-bus system --- Comparison of SE values obtained with expressions \eqref{eq:complex_TEO} and \eqref{eq:estimation_TEO}.}
    \label{fig:ieee14bus_stable_teager}
\end{figure}

The SE of machine 1 and the numerical estimation using~\eqref{eq:estimation_TEO} is depicted in Figure~\ref{fig:ieee14bus_stable_teager}. The SE converges to zero about 20 s, indicating that the machine 1 reaches a stationary operating condition and asymptotic local synchronization.  Moreover, we can note the consistency of the proposed complex-frequency-based formulation with the numerical-based estimation based on~\eqref{eq:estimation_TEO}. These results validate the complex frequency-based local synchronization conditions highlighted in \textit{Remark 2.1}.

This system is also used to assess the local strength of each machine's connection point, independent of its capacity.  Here, local strength refers to a node's ability to withstand variations in both angle and magnitude. To quantify this, we compute the normalized SE, obtained by normalizing \eqref{eq:final_TEO} by $2|\bar{s}_h|^2$. Based on this metric, we can identify which nodes are more susceptible to larger deviations between $\omega_v$ and $\omega_\imath$, as well as fluctuations in voltage and current magnitude.

\begin{figure}[htb!]
    \includegraphics[width=0.9\linewidth]{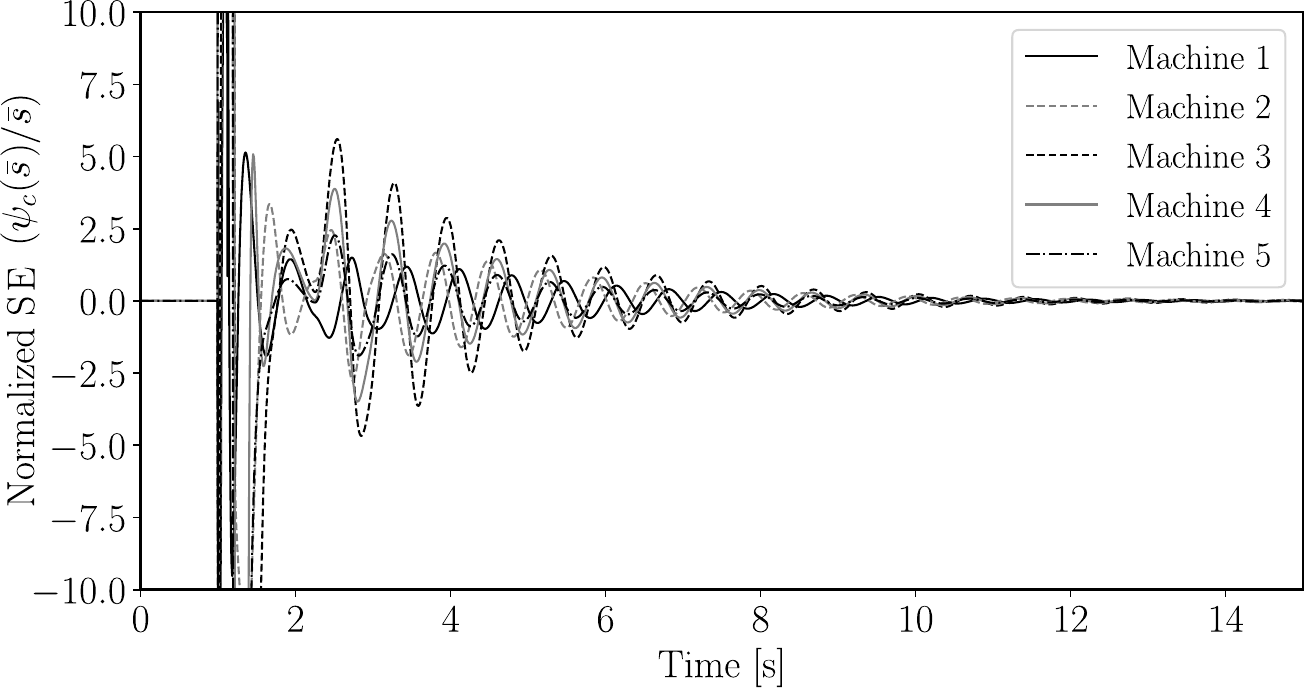}
    \centering
    \caption{IEEE 14-bus system --- Normalized SE of all machines for the scenario with stable steady-state operating condition.}
    \label{fig:normsync_stable}
\end{figure}

\begin{figure}[htb!]
    \includegraphics[width=\linewidth]{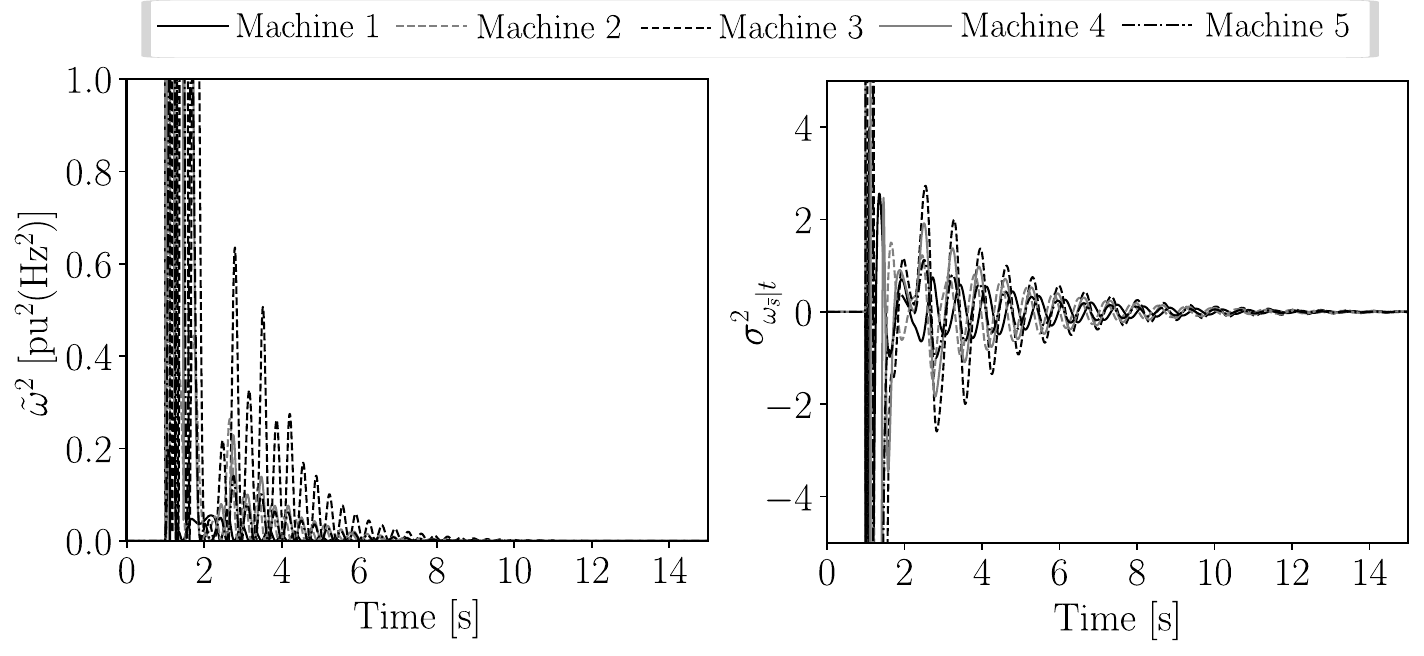}
    \centering
    \caption{IEEE 14-bus system --- Components of the SE for the scenario with stable steady-state operating condition.}
    \label{fig:key_stable}
\end{figure}

The normalized SE is calculated and presented in Figure~\ref{fig:normsync_stable}.  It can be observed that machine 3 is the most affected by variations in frequency ($\omega_v$ and $\omega_\imath$) and magnitude ($\rho_v$ and $\rho_{\imath}$) at its connection point.  These variations are further illustrated in Figure~\ref{fig:key_stable}, which shows $\Tilde{\omega}^2 = (\omega_v - \omega_\imath)^2$ and the conditional standard deviation of the complex power defined in~\eqref{eq:sigma2_sh}.

This result shows that although machine 1, the largest machine in the system, requires the most effort (in terms of SE) to remain synchronized after the disturbance, machine 3 is the one connected to the weakest node in the system.  To validate these results, we perform a conventional short-circuit analysis by calculating the Thevenin impedance and determining the short-circuit capacity (SCC) at each machine's connection bus.  The SCC is presented in Table~\ref{tab:short_circuit_capacity}. We can note that machine 3 is the one connected to the bus with the lower SCC. This results aligning with our findings based in the calculation of the normalized SE after a disturbance.

\begin{figure}[htb!]
    \includegraphics[width=0.9\linewidth]{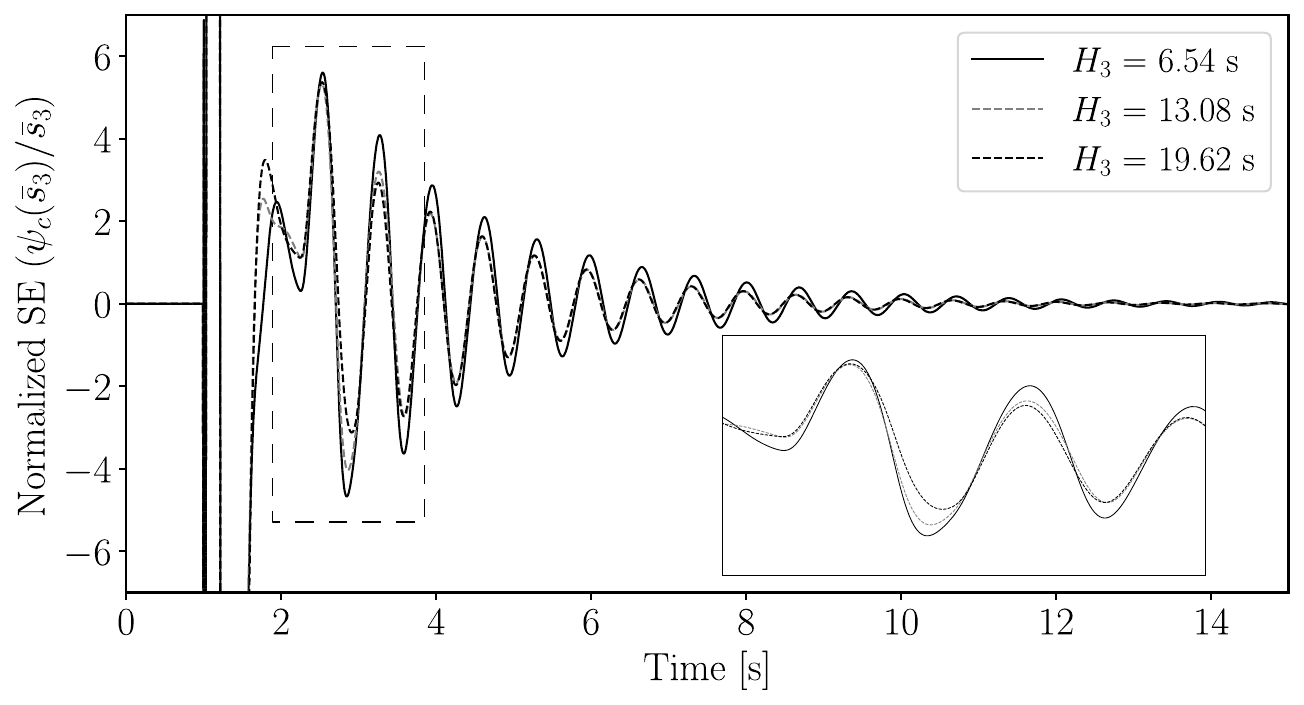}
    \centering
    \caption{IEEE 14-bus system --- Comparison of normalized SEs of machine 3 for three values of the inertia.}
    \label{fig:comparativeH_machine3}
\end{figure}

It is important to note that, unlike SE, SCC considers only the system topology. To illustrate this difference, we now consider three inertia cases of machine 3 (base, double, and triple) and carry out the simulations with the same contingency. The comparison is shown in Figure~\ref{fig:comparativeH_machine3}, where a clear reduction in normalized SE can be observed, an aspect that is not captured by the bus SCC.

\begin{table}[htb!]
\centering
\caption{IEEE 14-bus system --- Short Circuit Capacity of Connection Buses}
\begin{tabular}{cc}
\hline
\multirow{2}{*}{\centering Machine} & Short Circuit Capacity \\ 
 & (MVA) \\ \hline
1 & 297.1 \\ 
2 & 284.5 \\ 
3 & 270.7 \\ 
4 & 280.9 \\ 
5 & 317.3 \\ \hline
\end{tabular}
\label{tab:short_circuit_capacity}
\end{table}

\subsubsection{Stable limit cycle}

\begin{figure}[htb!]
    \includegraphics[width=\linewidth]{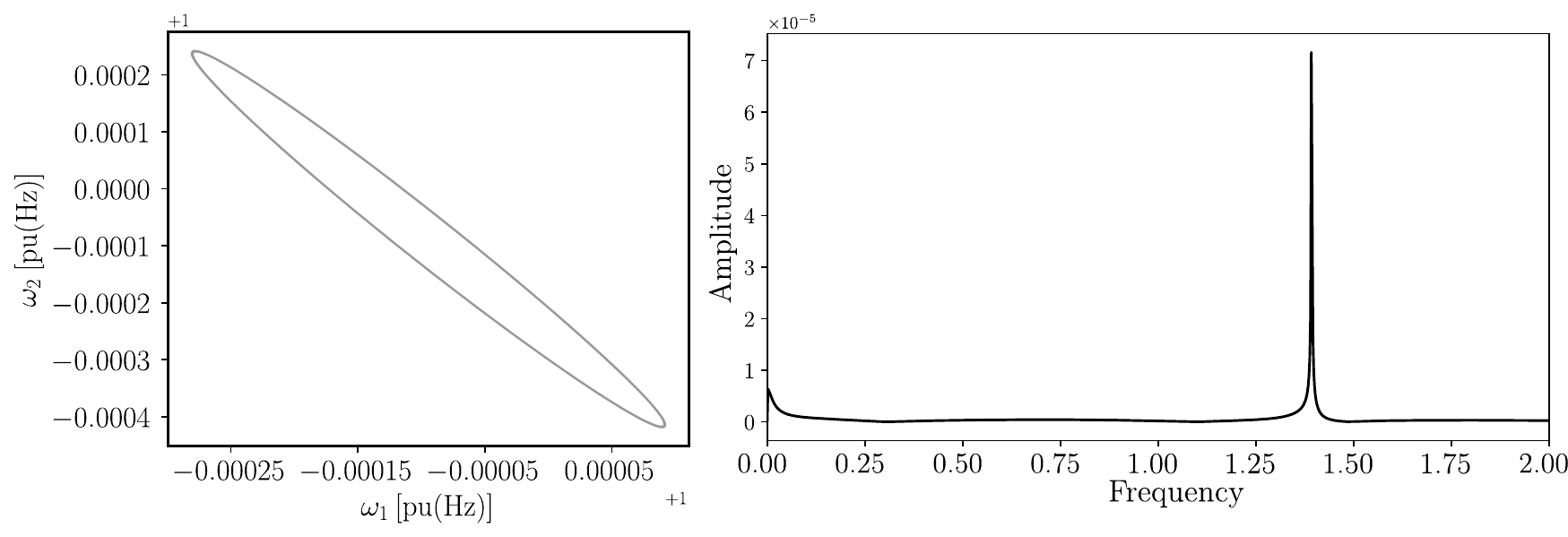}
    \centering
    \caption{IEEE 14-bus system --- Limit cycle and the spectrum of $\omega_1$.}
    \label{fig:ieee14bus_LC_w1w2}
\end{figure}

In this case, the demand is increased by 20\% and the same disturbance as in the previous case is applied.  Figure~\ref{fig:ieee14bus_LC_w1w2} shows the limit-cycle in the state space of the rotor speeds of machines 1 and 2, as well as the frequency spectrum obtained with the Fourier transform of the frequency of machine 1.  The oscillation period is approximately 0.7 s.  Note that the stability of the limit cycle can be confirmed with the Floquet multipliers of the monodromy matrix of the system's trajectory~\cite{Bizzarri2016}.

\begin{figure}[htb!]
    \includegraphics[width=0.9\linewidth]{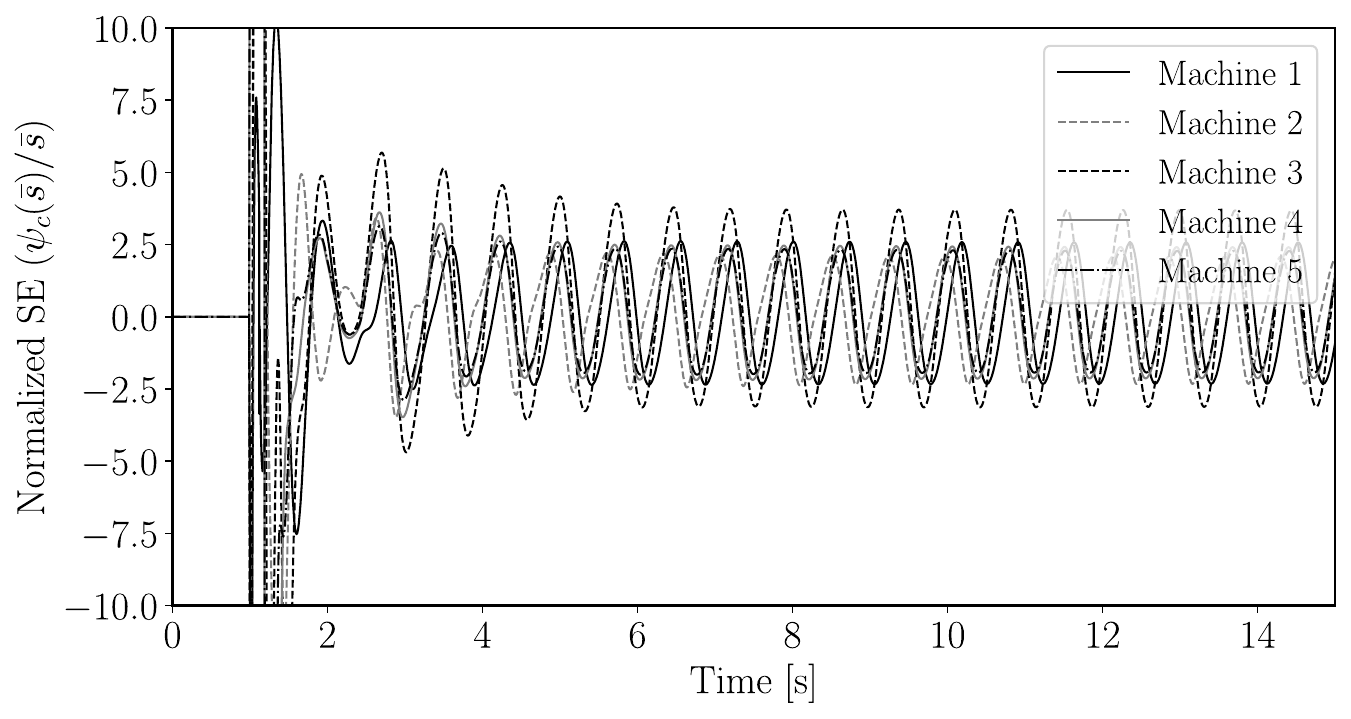}
    \centering
    \caption{IEEE 14-bus system --- Normalized SEs of synchronous machines for the case that leads to a stable limit cycle.}
    \label{fig:ieee14bus_LC_teager}
\end{figure}

The resulting normalized SE following the transient behavior is shown in Figure \ref{fig:ieee14bus_LC_teager}.  As expected, machine 3 exhibits the highest synchronization energy. In this case the machines do not achieve asymptotic local synchronism, despite the system being stable.  

\begin{figure}[htb!]
    \includegraphics[width=0.9\linewidth]{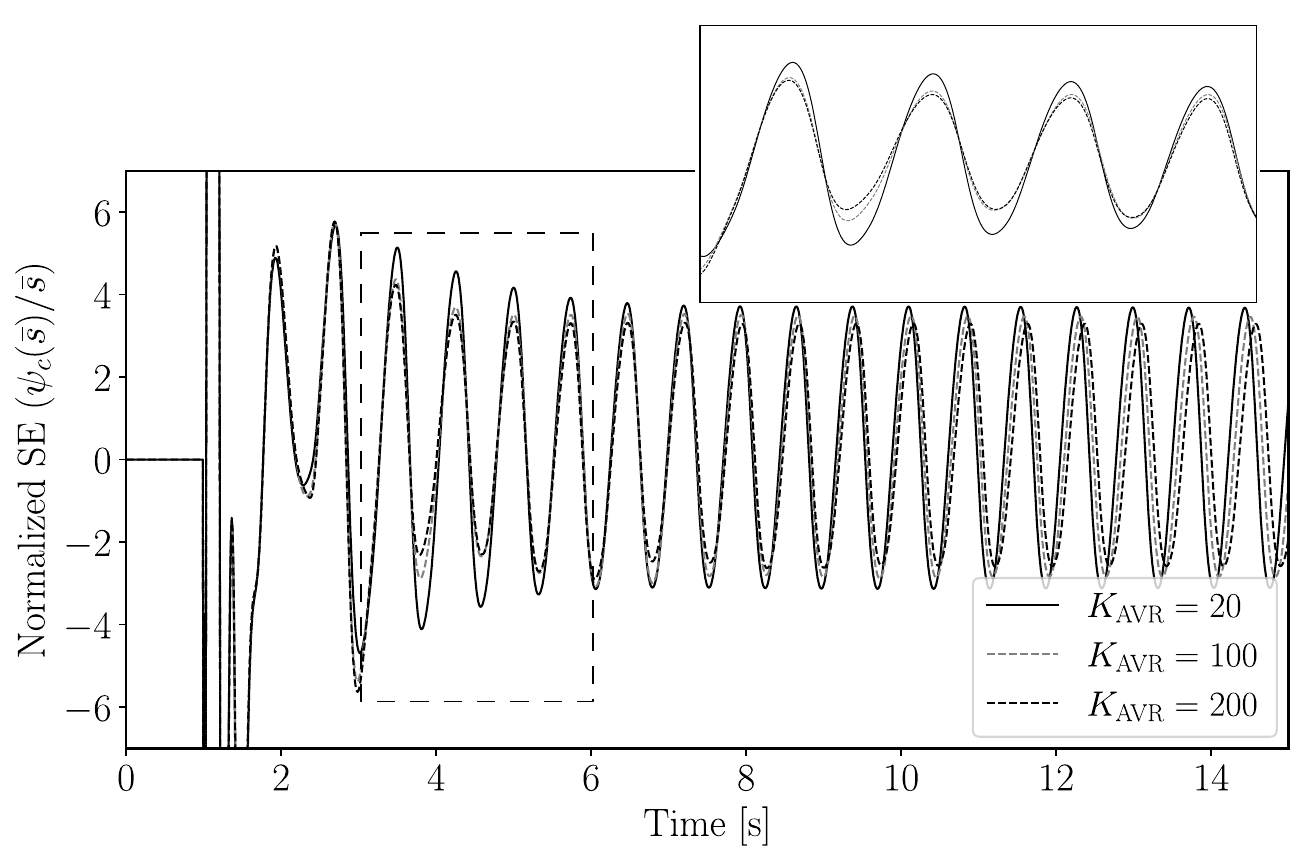}
    \centering
    \caption{IEEE 14-bus system --- Normalized SEs of machine 3 for different AVR gains for the case that leads to a stable limit cycle.}
    \label{fig:ieee14bus_LC_teager_AVR}
\end{figure}

Next, we show that the SE captures the impact of control parameters on local synchronization. With this aim, we evaluate three cases with different AVR gain ($K_{\rm AVR}$) values for machine 3.  The resulting normalized synchronization energy is depicted in Figure~\ref{fig:ieee14bus_LC_teager_AVR}.  In this specific case, increasing the AVR gain of the machine contributes to a more robust bus connection, that is, a reduction in the normalized SE of machine 3. 

Therefore, synchronism and system strength are expected to be closely linked, i.e., for a device to remain synchronized for different contingencies, it should be connected to a robust system. The results for the IEEE 14-bus system highlight the potential of normalized SE in assessing system strength, or, using the definition proposed in \cite{Aleksandar2022}, dynamic-state system strength. This aspect will be explored in future work.

\subsection{Grid Following - IBR system}
\label{sub:ibr}

\begin{figure}[htb!]
    \includegraphics[width=0.9\linewidth]{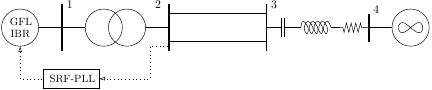}
    \centering
    \caption{Single line diagram of GFL-IBR with series compensation.}
    \label{fig:gfl}
\end{figure}

This case involves a grid-following inverter-based resource (GFL IBR) and a series compensation, as depicted in the line diagram of Figure~\ref{fig:gfl}.  The GFL model are with a PI controller for the internal current control loops and with a synchronous reference Phase Locked Loop (SRF-PLL) for synchronization with the system.  The transmission lines, transformer and series compensation are modeled based on Park vectors retaining the electromagnetic dynamics.  Further details regarding the GFL model can be found in \cite{ponce2024localsync}.  The SRF-PLL is controlled by a PI controller.  In this study, we fix the proportional gain at 0.28 and examine two cases with different integrator gains ($K_\imath$). 

\begin{figure}[htb!]
    \includegraphics[width=0.9\linewidth]{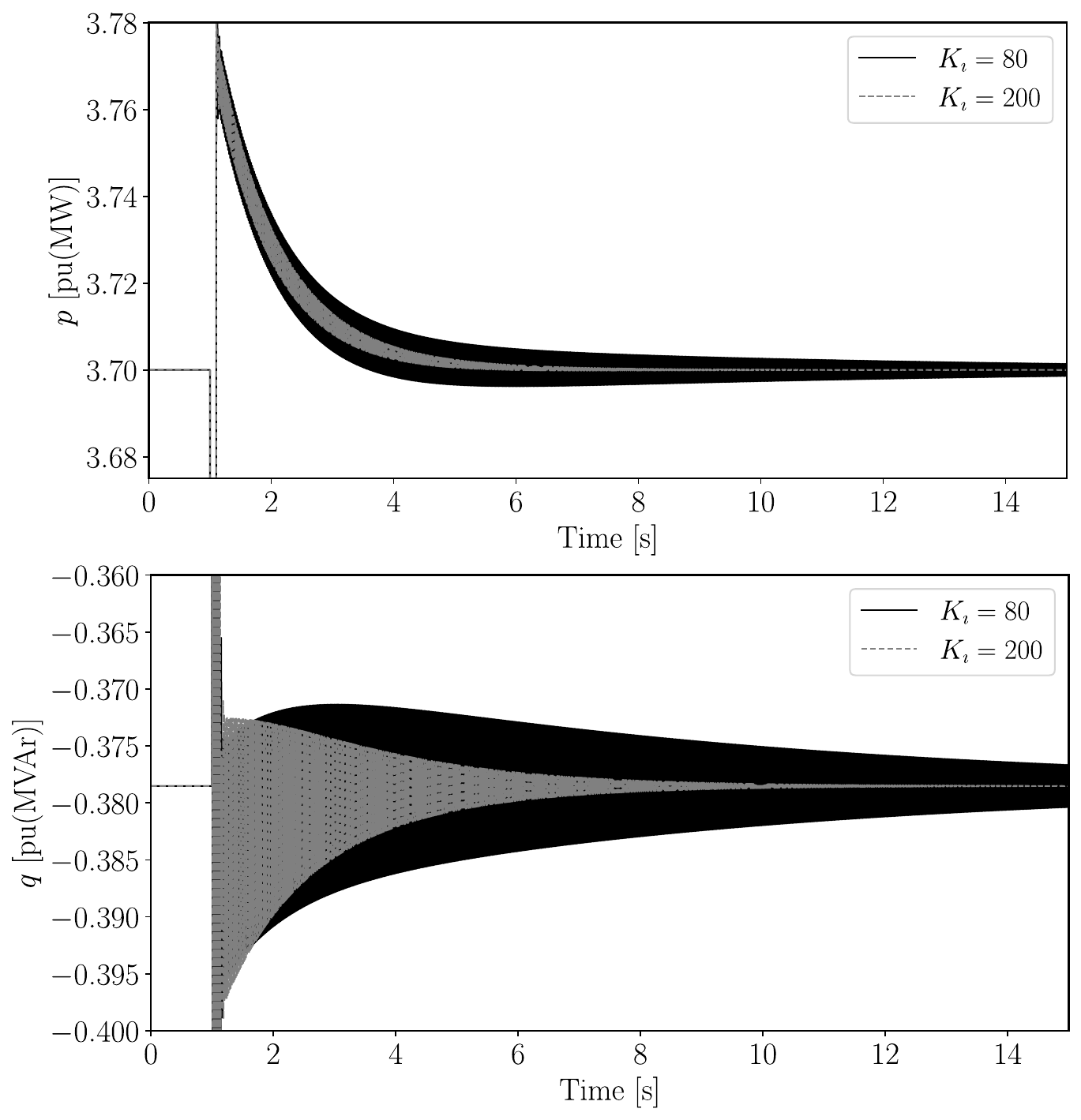}
    \centering
    \caption{GFL system --- Active and reactive powers injected by the GFL-IBR.}
    \label{fig:gfl_powers}
\end{figure}

\begin{figure}[htb!]
    \includegraphics[width=0.9\linewidth]{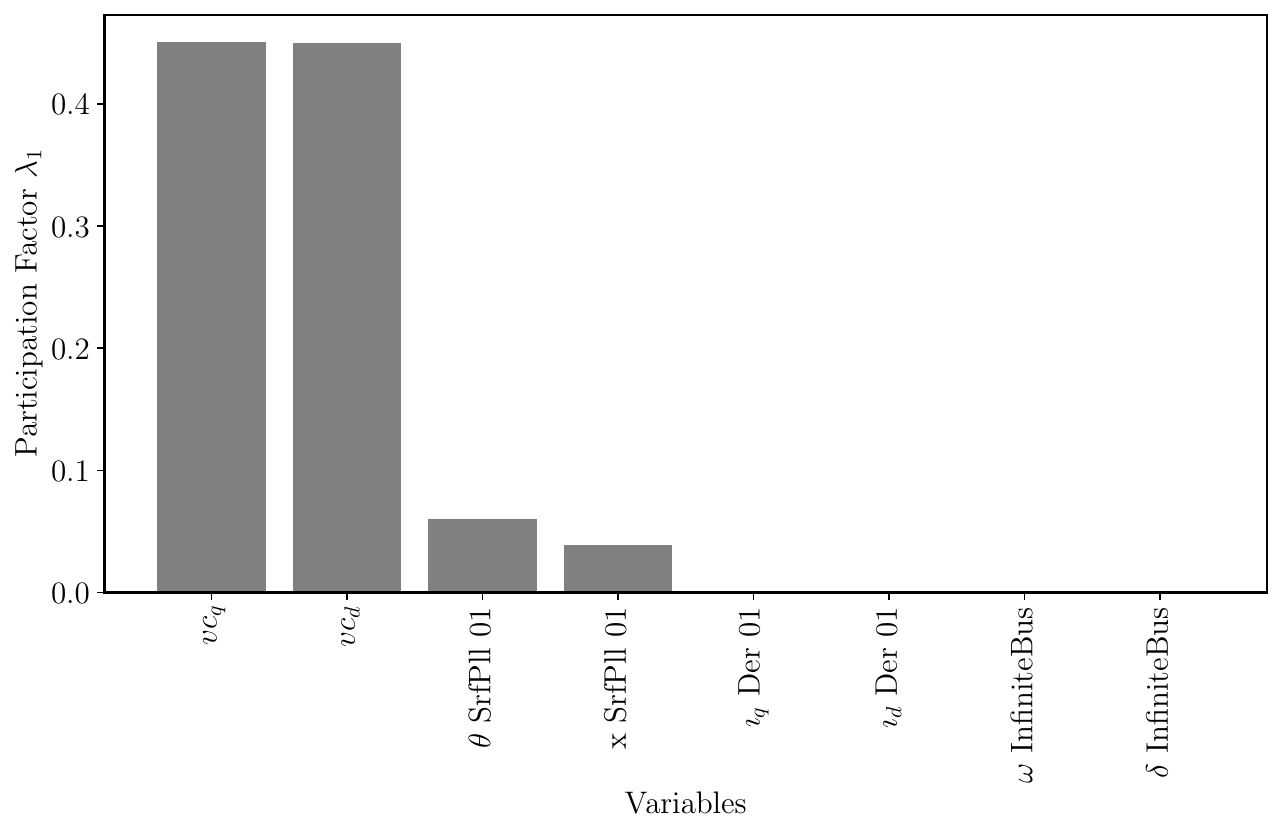}
    \centering
    \caption{Participation factors of the states variables of the GFL-IBR system for the undamped oscillation mode (case with $K_i = 80$).}
    \label{fig:pfactors}
\end{figure}

The disturbance consists of a short-circuit applied at $t = 1$ s on the GFL connection bus, which is cleared at $t=1.1$ s by opening one of the line circuits. Figure~\ref{fig:gfl_powers} shows the complex power for the two cases: $K_\imath = 80$, and $K_\imath = 200$.  In both cases the GFL-IBR are in resonance with the series compensation, leading to high-frequency oscillations following the disturbance. To further analyze this behavior, we perform a small-signal analysis for the case with $K_\imath = 80$. The undamped oscillation mode is $0.119 \pm j376.8$, corresponding to a natural frequency of 59.97 Hz.  The participation factors of the state variables associated with this oscillation mode are shown in Figure \ref{fig:pfactors}. Notably, there is a high participation of the series compensator voltage ($vc_d$ and $vc_d$) 
and the states of the SRF-PLL ($\theta$ and $x$, where $x$ is the internal state of the PI control).

\begin{figure}[htb!]
    \includegraphics[width=0.9\linewidth]{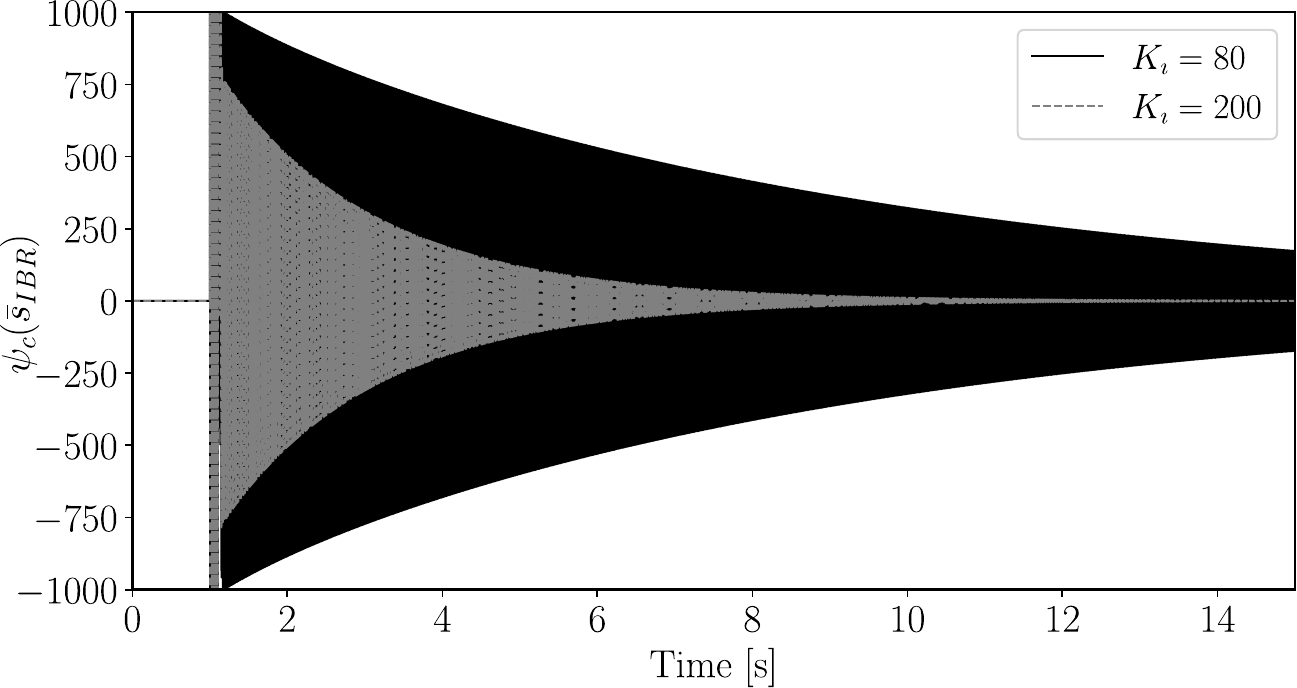}
    \centering
    \caption{SE of the GFL-IBR for different SRF-PLL gains.}
    \label{fig:gfl_gains}
\end{figure}

The resulting synchronization energy for both cases is illustrated in Figure~\ref{fig:gfl_gains}. In both scenarios, the GFL-IBR attains asymptotic local synchronization, despite exhibiting high synchronization energy levels.  Notably, the case with the higher integrator gain leads to faster convergence to asymptotic local synchronism and lower SE.  These results demonstrating the impact of tuning the PI controller on the system's dynamic response and, consequently in the SE.

\section{Conclusions}
\label{sec:conclusion}

This work introduces an application of the TEO to the complex power injected at power system buses and proposes a novel definition of \textit{synchronization energy} (SE).  This quantity can be interpreted as a metric to account for the dynamic behavior of both active and reactive power, providing a more comprehensive and synthetic assessment of local synchronization.  We demonstrates that these variations are intrinsically linked to the difference between the instantaneous frequencies of voltage and current, as well as the magnitudes of voltage at the connection bus and the device’s current magnitude.  Through a comprehensive set of case studies, we show that the proposed SE effectively captures the impact of key internal device parameters, such as inertia of synchronous machines, transmission system reactances, and controller gains of IBRs.  We believe that this wide sensitivity range makes the proposed SE a useful tool for evaluating the dynamic performance of modern power systems. 

Future research will explore the application of SE in control strategies and exploring its potential in identifying system strength, offering a promising direction for improving the stability and reliability of power systems.

\section*{Acknowledgments}

This work was supported in part by Brazilian Agency Coordination for the Improvement of Higher Education Personnel (CAPES) (Finance Code 001) and by Engie under the Research and Development Program regulated by the Brazilian Electricity Regulatory Agency (ANEEL) (PD-00403-0053/202), by funding B.~Pinherio; and by Sustainable Energy Authority of Ireland (SEAI) by funding I.~Ponce and F.~Milano under project FRESLIPS, Grant No.~RDD/00681.

\appendix

\section{Wigner Distribution}

The Wigner distribution represents a signal in both time and frequency domains simultaneously, offering a detailed view of how the frequency content of a signal evolves over time \cite{LeonCohen}. 

Given a complex signal $\bar{x}(t) = a(t)e^{\phi(t)}$, the Wigner distribution $W(t,f)$ is defined as follows:
\begin{equation}
    W(t,\omega) = \frac{1}{2\pi} \int \bar{x}^*(t - \frac{1}{2}\tau) \bar{x}(t + \frac{1}{2}\tau)e^{-j\tau\omega} d\tau ,
\end{equation}
where $t$ is time, $\omega$ is frequency, $\tau$ is a time-shift, and $\bar{x}^*(t)$ is the complex conjugate of $\bar{x}(t)$. 

The first conditional moment in frequency of the Wigner distribution is given by
\begin{equation}
    \label{eq:conditional_1}
    \langle \omega \rangle_t = \frac{1}{|\bar{x}(t)|^2} \int \omega W(t,\omega) d\omega = \dot\phi(t),
\end{equation}
that is, the average frequency at a given time $t$ is the instantaneous frequency of $\bar{x}(t)$.  The second conditional moment in frequency is defined as~\cite{LeonCohen}:
\begin{align}
    \label{eq:conditional_2}
   \langle \omega^2 \rangle_t &= \frac{1}{|\bar{x}(t)|^2} \int \omega^2 W(t,\omega) d\omega \\
   &= \frac{1}{2}\left[ \left( \frac{ \dot a}{a} \right)^2 - \frac{\ddot a}{a} \right] + \dot \phi^2(t) .
\end{align}
Then, the conditional standard deviation in frequency of the signal $\bar{x}(t)$ is given by:
\begin{align}
    \label{eq:conditional_3}
    \sigma^2_{\omega_x|t} &= \langle \omega^2 \rangle_t- \langle \omega \rangle_t^2 = \frac{1}{2}\left[ \left( \frac{ \dot a}{a} \right)^2 - \frac{\ddot a}{a} \right] .
\end{align}

The conditional standard deviation is the spread of the frequency about the instantaneous frequency.  This quantity can be negative, which makes problematic its physical interpretation.

%\bibliographystyle{bibliography/IEEEtran}
%\bibliography{bibliography/IEEEabrv,./bibliography/refs}


\end{document}